\documentclass[apj]{emulateapj}

\usepackage{amsmath}
\shortauthors{LAPI ET AL.}
\shorttitle{PRECISION SCALING RELATIONS FOR DISK GALAXIES}
\slugcomment{ACCEPTED BY ApJ}

\begin{document}

\title{Precision Scaling Relations for Disk Galaxies in the Local Universe}
\author{A. Lapi\altaffilmark{1,2,3}, P. Salucci\altaffilmark{1,2,3}, L. Danese\altaffilmark{1,3}}
\altaffiltext{1}{SISSA, Via Bonomea 265, 34136 Trieste, Italy}\altaffiltext{2}{INFN-Sezione di Trieste, via Valerio 2, 34127 Trieste,  Italy}\altaffiltext{3}{INAF-Osservatorio Astronomico di Trieste, via Tiepolo 11, 34131 Trieste, Italy}

\begin{abstract}
We build templates of rotation curves as a function of the $I-$band luminosity via the mass modeling (by the sum of a thin exponential disk and a cored halo profile) of suitably normalized, stacked data from wide samples of local spiral galaxies. We then exploit such templates to determine fundamental stellar and halo properties for a sample of about $550$ local disk-dominated galaxies with high-quality measurements of the optical radius $R_{\rm opt}$ and of the corresponding rotation velocity $V_{\rm opt}$. Specifically, we determine the stellar $M_\star$ and halo $M_{\rm H}$ masses, the halo size $R_{\rm H}$ and velocity scale ${V_{\rm H}}$, and the specific angular momenta of the stellar $j_\star$ and dark matter $j_{\rm H}$ components. We derive global scaling relationships involving such stellar and halo properties both for the individual galaxies in our sample and for their mean within bins; the latter are found to be in pleasing agreement with previous determinations by independent methods (e.g., abundance matching techniques, weak lensing observations, and individual rotation curve modeling). Remarkably, the size of our sample and the robustness of our statistical approach allow us to attain an unprecedented level of precision over an extended range of mass and velocity scales, with $1\sigma$ dispersion around the mean relationships of less than $0.1$ dex. We thus set new standard local relationships that must be reproduced by detailed physical models, that offer a basis for improving the sub-grid recipes in numerical simulations, that provide a benchmark to gauge independent observations and check for systematics, and that constitute a basic step toward the future exploitation of the spiral galaxy population as a cosmological probe.
\end{abstract}


\setcounter{footnote}{0}

\section{Introduction}\label{sec|intro}

In recent years the fields of cosmology and of galaxy formation have become more and more interlinked: precision cosmology demands for a careful control of systematics, whose origin is in many contexts related to the astrophysics of galaxies; on the other hand, a galaxy population with homogeneous properties and controlled evolution can be exploited as tracer of the underlying matter field and hence as a cosmological probe. In this vein, a relevant role will be played by spiral, disk-dominated galaxies forming and evolving at $z\la 1.5$, a crucial period where the dark energy kicks in and enforces the accelerated cosmic expansion.

A fundamental step toward exploiting spiral galaxies for cosmological purposes is to characterize the properties of their baryonic and dark matter (DM) components. Specifically, it is of paramount importance to accurately determine the mass of the stars $M_\star$ and of the host DM halo $M_{\rm H}$, the scales $R_{\rm opt}$\footnote{The size of the stellar component can be equivalently characterized by the exponential disk scale-length $R_d$, by the effective (half-mass) radius $R_e\approx 1.68\, R_d$, and by the optical radius $R_{\rm opt}\equiv 3.2\, R_d$.} and $R_{\rm H}$ over which stars and DM are distributed, the typical velocities $V_{\rm opt}$ and $V_{\rm H}$ measuring their contribution to the total gravitational potential, and the associated specific angular momenta $j_\star$ and $j_{\rm H}$. In the past literature such a challenging task has been addressed via a variety of methods (see Courteau et al. 2014 for a review), including satellite kinematics (see More et al. 2011; Wojtak \& Mamon 2013) weak lensing observations (see Reyes et al. 2012; Velander et al. 2014; Hudson et al. 2015; Mandelbaum et al. 2016), (subhalo) abundance matching (see Shankar et al. 2006; Moster et al. 2013; Aversa et al. 2015; Rodriguez-Puebla et al. 2015; Huang et al. 2017), and individual rotation curve (RC) modeling (see Romanowsky \& Fall 2012; Lelli et al. 2016). However, previous studies have relied on samples with limited statistics, have investigated a narrow range of masses/spatial scales/kinematic properties, and have focused on a specific aspect (e.g., the $M_\star-M_{\rm H}$ relationship).

In the present paper we attack the issue with an alternative statistical approach (e.g., Persic et al. 1996; Catinella et al. 2006: Salucci et al. 2007; Yegorova et al. 2011). First, we build RC templates as a function of $I-$band luminosity via the mass modeling (by the sum of a thin exponential disk and a cored halo profile) of suitably normalized, stacked data from wide samples of local spiral galaxies. We then exploit such templates to determine fundamental stellar and halo properties for a sample of about $550$ local disk-dominated galaxies with high-quality measurements of the optical radius $R_{\rm opt}$ and of the corresponding rotation velocity $V_{\rm opt}$. Specifically, we determine the stellar $M_\star$ and halo $M_{\rm H}$ masses, the halo size $R_{\rm H}$ and velocity scale $V_{\rm H}$, and the specific angular momenta of the stellar $j_\star$ and DM $j_{\rm H}$ components. Finally, we combine these stellar and DM quantities to derive high-precision global scaling relationships, so providing a new benchmark for understanding the astrophysics of spirals and at the same time paving the way for their future cosmological exploitation.

The plan of the paper is as follows: in Sect.~\ref{sec|data} we describe in some details our analysis to derive the templates, and their exploitation to determine global quantities characterizing the DM and stellar components for a specific sample of about $550$ local disks; in Sect.~\ref{sec|results} we present and discuss our results, confronting them with independent datasets and with the outcomes expected from the classic theory of disk formation; in Sect.~\ref{sec|summary} we summarize our findings and provide future perspectives on their use for cosmological studies.

Throughout this work, we adopt the standard flat cosmology (Planck Collaboration XIII 2016) with round parameter values: matter density $\Omega_{\rm M}=0.32$, baryon density $\Omega_b = 0.05$, Hubble constant $H_0 = 100\, h$ km s$^{-1}$ Mpc$^{-1}$ with $h=0.67$, and mass variance $\sigma_8 = 0.83$ on a scale of $8\, h^{-1}$ Mpc. Stellar masses and SFRs (or luminosities) of galaxies are evaluated assuming the Chabrier (2003) initial mass function.

\section{Data analysis}\label{sec|data}

Our approach to characterize the stellar and dark matter properties of local spiral galaxies consists of two steps: (i) the construction of RC templates; (ii) their exploitation to infer global properties for a specific sample of local disk-dominated galaxies. We now describe these in turn.

\subsection{Modeling of stacked rotation curves}\label{sec|URC}

We build up RC templates for galaxies of different luminosities by exploiting the Universal Rotation Curve (URC) statistical approach (see Persic et al. 1996; Salucci \& Burkert 2000; Salucci et al. 2007). This is based on the notion, implicit even in the seminal paper by Rubin et al. (1985), that the RCs of local spiral galaxies with given luminosity are remarkably similar in shape. Then the mass modeling is performed not on the individual RCs, but on the stacked, suitably normalized RCs of galaxies within a given luminosity bin.

Such a statistical approach has some relevant advantages over the brute force fit to individual RCs: (i) it increases the signal-to-noise allowing for a precise description of the average RCs, smoothing out small-scale fluctuations induced by bad data points and/or by physical features such as spiral warps; (ii) it allows the simultaneous mass decomposition of RCs for galaxies with similar luminosity but with different spatial sampling.

Typically, the stacked data are constructed as follows: the individual RCs $V(R)$ are normalized in both the dependent and independent variables in terms of the optical radius $R_{\rm opt}$ and of the corresponding velocity $V_{\rm opt}\equiv V(R_{\rm opt})$, to obtain a normalized curve $\tilde V(\tilde R)\equiv V(R/R_{\rm opt})/V_{\rm opt}$; then all the normalized RCs of individual galaxies falling in the same magnitude bin are coadded (properly weighting measurement uncertainties), to derive a stacked curve $\tilde V(\tilde R|M_I)$. Note that the choices of $R_{\rm opt}$ and $V_{\rm opt}$ as normalization points are meant to minimize the uncertainties in velocity estimates because there the RC is flattening and the disk surface brightness is quickly decreasing.

We base our analysis on three stacked RC compilations from the literature. The first one has been constructed by Persic et al. (1996) from about $900$ individual RCs extending out to $R\la R_{\rm opt}$, separated into $11$ luminosity bins with average $I-$band magnitude from $\langle M_I\rangle\sim -18.5$ to $-23.2$. The second stacked curve compilation has been constructed by Catinella et al. (2006) from a sample of about $2200$ individual RCs extending out to $1.5-2\,R_{\rm opt}$, separated into $10$ $I-$band magnitude bins with average values from $\langle M_I\rangle\sim -19.4$ to $-23.8$. The third stacked curve compilation has been constructed by Yegorova et al. (2011) from a sample of $30$ individual RCs extending out to $\la 2\, R_{\rm opt}$ for high-luminosity spirals with average $I-$band magnitude $\langle M_I\rangle\sim -23.3$. Uncertainties around the mean for the stacked normalized curves are of order $\la 5\%$, typically increasing toward the outer regions.

We analyse these stacked RCs with the following physical template
\begin{equation}\label{eq|URC}
\tilde V_{\rm TOT}^2(\tilde R|M_I) = \tilde V_{\rm DISK}^2(\tilde R|M_I)+\tilde V_{\rm HALO}^2(\tilde R|M_I)~,
\end{equation}
which is an addition of two terms: (i) an exponential, infinitesimally thin disk profile (Freeman 1970)
\begin{equation}\label{eq|URC_disk}
\tilde V_{\rm DISK}^2(\tilde R|M_I) = \kappa\, \tilde R^2\, {B(1.6\tilde R)\over B(1.6)}
\end{equation}
where $B(x)\equiv I_0(x)\, K_0(x)-I_1(x)\, K_1(x)$ is a combination of modified Bessel functions (for finite thickness the rotation curve will have a very similar shape but for a $5\%$ lower peak, see Casertano 1983); (ii) a cored dark halo profile following the Burkert shape (see Burkert 1995; Katz et al. 2017)
\begin{equation}\label{eq|URC_halo}
\tilde V_{\rm HALO}^2(\tilde R|M_I) = (1-\kappa)\, {1\over \tilde R}\,{G(\tilde R/\tilde R_c)\over G(1/\tilde R_c)}~
\end{equation}
where $G(x)=\ln(1+x)-{\rm tan}^{-1}(x)+0.5\,\ln(1+x^2)$. This mass decomposition involves two parameters, namely the normalized core radius $\tilde R_c(M_I)$ and the relative contribution $\kappa(M_I)$ of the disk to the overall template at the optical radius. The main assumption in the construction of a universal template is that these parameters assume a single value within a (small) magnitude bin; this implies (see next Section) that the stellar $M_\star\propto \kappa\, V_{\rm opt}^2\, R_{\rm opt}$ and halo masses $M_{\rm H}\propto (1-\kappa)\, V_{\rm opt}^2\, R_{\rm opt}/G(R_{\rm opt}/R_c)$ of each galaxy within a magnitude bin are determined only by further specifying the observed optical radius $R_{\rm opt}$ and velocity $V_{\rm opt}$.

In the Appendix we discuss to what extent the results of our analysis are affected by adopting shapes of the DM halo different from Burkert's, such as the classic NFW profile extracted from $N-$body simulations (e.g., Navarro et al. 1997, 2010; Ludlow et al. 2013; Peirani et al. 2017), and the profile indicated by hydrodynamical simulations including baryonic effects on the DM halo (e.g., Di Cintio et al. 2014). In a nutshell, cuspier halo density
profiles in the inner region yield a lower disk contribution, hence
a smaller stellar mass. The inferred halo mass is to some extent
less sensitive to the halo profile, since for the majority of galaxies depends on the outer behavior of the RC, which is similar. However, in faint galaxies where the disk contribution is marginal or negligible, the inferred halo mass is also sensitive to the inner shape of the RCs, and it is found to be larger for cuspier density profiles. All that adds to the well-known diversity in the fitting accuracy of the RCs for strictly dwarf galaxies (not included in our sample), which appreciably increases when cored profiles are exploited (e.g., Gentile et al. 2004; Karukes \& Salucci 2017).

\begin{figure*}
\epsscale{0.9}\plotone{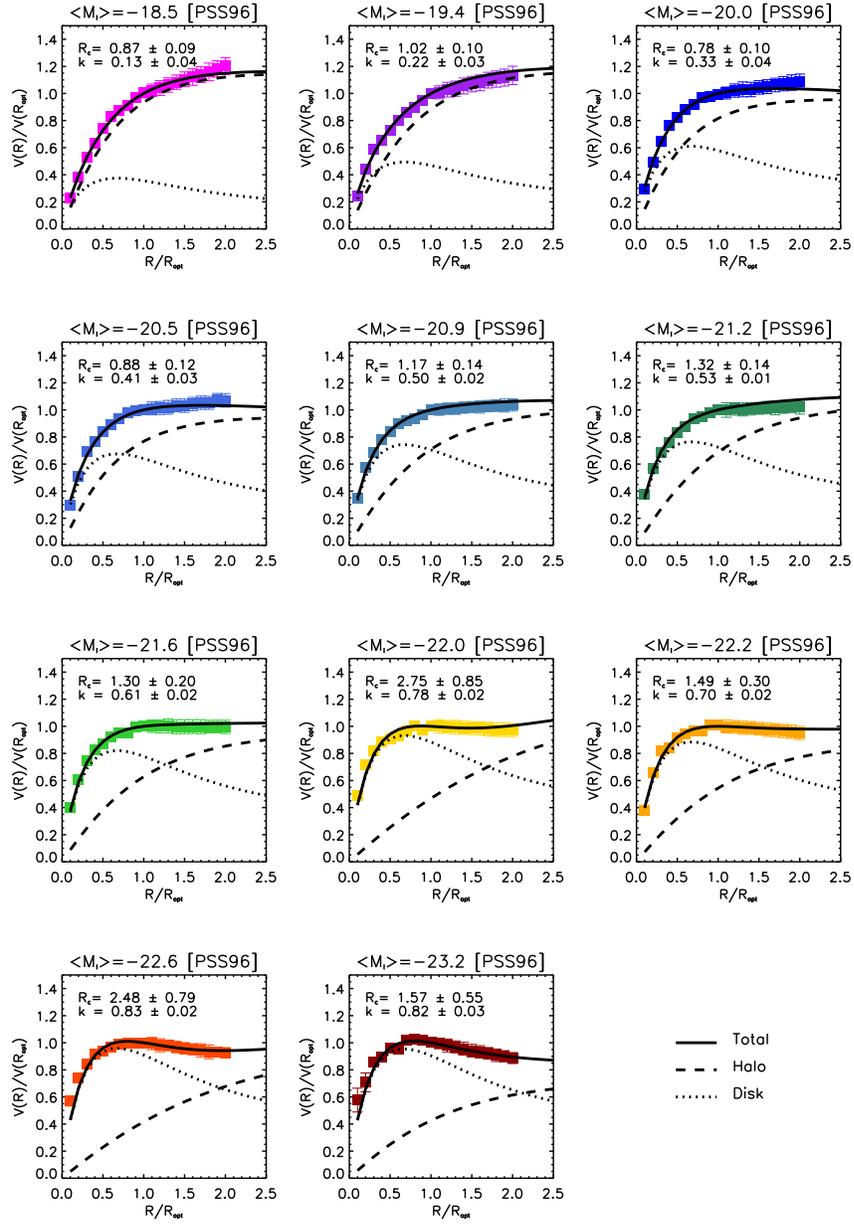}\caption{Modeling of the normalized stacked RCs from Persic et al. (1996), subdivided in $11$ $I-$band magnitude bins (average value of $\langle M_I\rangle$ as labeled). The overall RC template is illustrated as a solid line, the disk contribution as a dotted line and the halo contribution as a dashed line (see Sect.~\ref{sec|URC} for details). In each panel the best-fit parameters $\tilde R_c$ and $\kappa$ of the fit are also reported.}\label{fig|URC_fit}
\end{figure*}

\begin{figure*}
\epsscale{0.9}\plotone{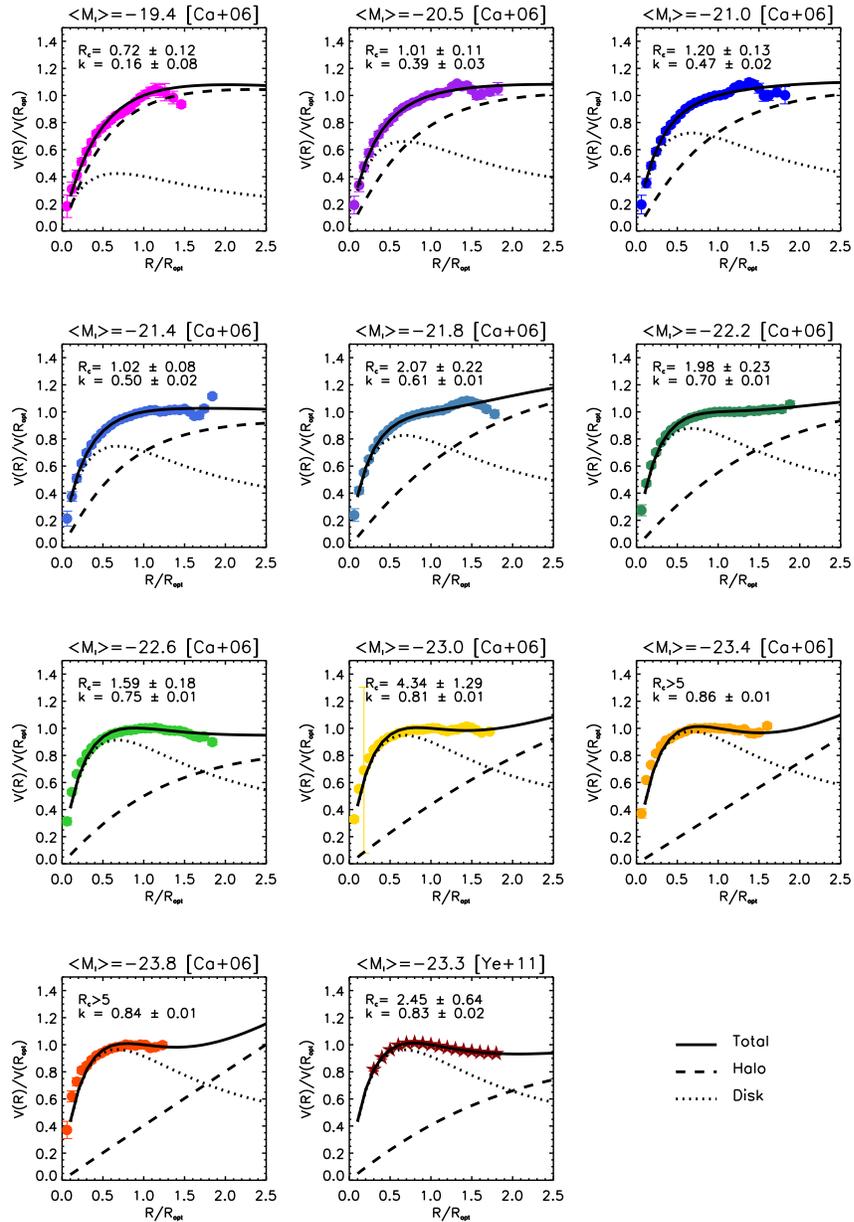}\caption{Same as previous figure for the stacked RCs from Catinella et al. (2006) in $10$ $I-$band magnitude bins (average value $\langle M_I\rangle$ as labeled) and for the high-luminosity disks from Yegorova et al. (2011). Linestyles as in previous Figure. }\label{fig|URC_fit_bis}
\end{figure*}

For each of the three samples described above and for any of the respective magnitude bins, we fit the stacked RCs with Eqs.~(\ref{eq|URC}-\ref{eq|URC_disk}-\ref{eq|URC_halo}) and determine the best fit parameters $\kappa$ and $\tilde R_c$; the results of the fitting to the stacked RCs with our templates are reported in Fig.~\ref{fig|URC_fit} and \ref{fig|URC_fit_bis}. The overall fits are reasonably good, albeit for some high-luminosity bins the shape of the outer RCs is not very well constrained and this causes large uncertainties on the estimated $\tilde R_c$. In these respects, the sample by Yegorova et al. (2011) provides the best constraints for high-luminosity disks, suggesting rather flat shapes beyond $R_{\rm opt}$. This is also confirmed from recent, high-resolution radio observations of a few RCs out to $\la 3\, R_{\rm opt}$ by Martinsson et al. (2016).
RC measurements of such a wide spatial extension in an ensemble of local bright spirals would be extremely relevant to improve the robustness of the stacked template at high luminosity.

\begin{figure*}
\epsscale{0.8}\plotone{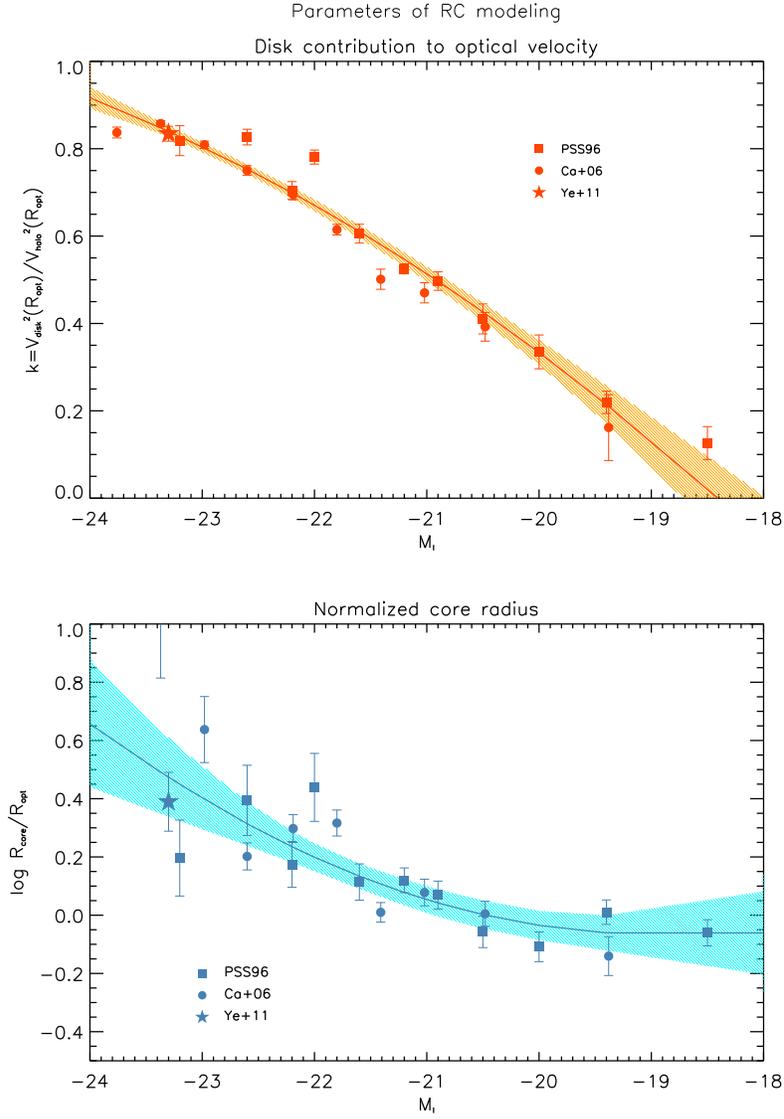}\caption{Best-fit parameters $\kappa$ (top panel) and $\tilde R_c$ (bottom panel) from the mass modeling of the stacked RCs from Persic et al. (1996; squares), Catinella et al. (2006; circles) and Yegorova et al. (2011; stars). In both panels the solid line is the analytical rendition (with the shaded area illustrating the associated $1\sigma$ uncertainty) in terms of Eq.~(\ref{eq|URC_parfit}).}\label{fig|URC_parfit}
\end{figure*}

The values and uncertainties of the best-fit parameters for the three different samples are reported in Fig.~\ref{fig|URC_parfit}. We find that the luminosity dependence of the parameters is rather weak, and can be rendered with the approximated expressions
\begin{eqnarray}\label{eq|URC_parfit}
\nonumber \log \tilde R_c(M_I) &\simeq & 0.182+0.421\,\log(\tilde L_I)+0.178\,[\log(\tilde L_I)]^2\\
\\
\nonumber \kappa(M_I) &\simeq& 0.656+0.369\,\log(\tilde L_I)-0.072\,[\log(\tilde L_I)]^2
\end{eqnarray}
where $\tilde L_I = 10^{-(M_I+21.9)/2.5}$; these are also illustrated, together with the associated uncertainties, in Fig.~\ref{fig|URC_parfit}.

\begin{figure*}
\epsscale{1}\plotone{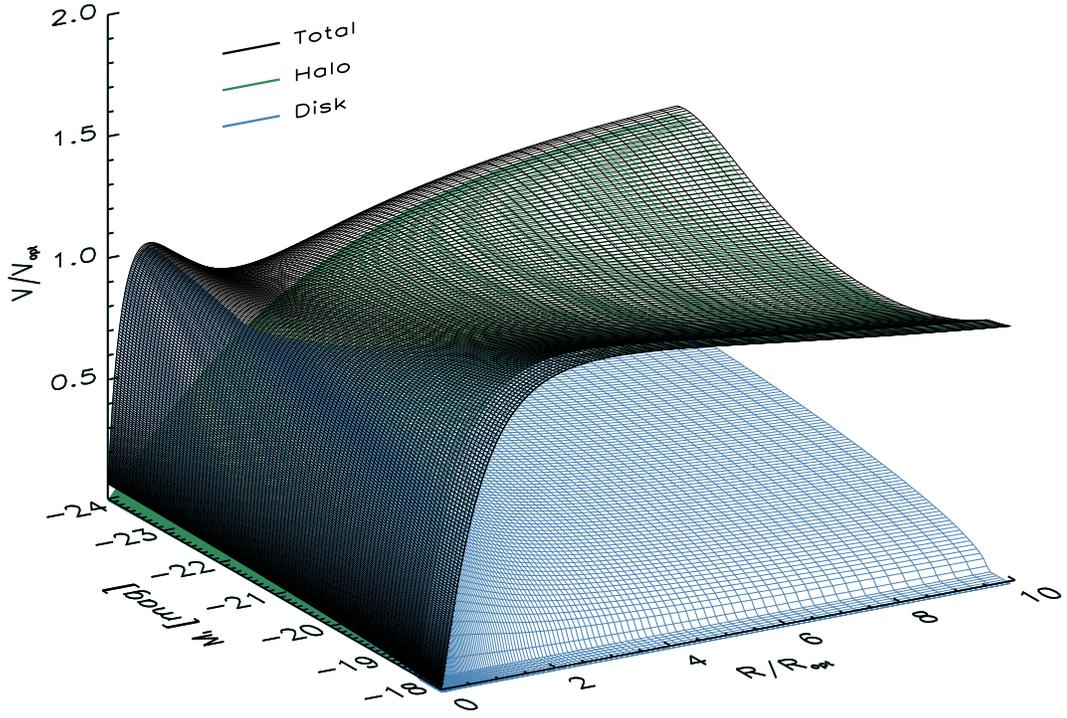}\caption{Our template $\tilde V_{\rm TOT}(\tilde R|M_I)$ is plotted in normalized units of $V_{\rm opt}$ against the radius $\tilde R=R/R_{\rm opt}$ normalized to $R_{\rm opt}$ and against the $I-$band magnitude $M_I$. Black surface refers to the total template, green to the halo component, and blue to the disk component.}\label{fig|URC}
\end{figure*}

The resulting templates as a function of radius and $I-$band magnitude are plotted in Fig.~\ref{fig|URC}, with the contribution of the disk and of the halo highlighted. As expected, it is seen that in moving toward brighter galaxies the contribution of the disk in the inner regions becomes predominant, while the rising portion of the RC enforced by the dark halo extends to outer and outer radii. We stress that the above method does not require to adopt a specific value for the mass-to-light ratio $M_\star/L$, but only to assume its dependence with radius (see Portinari \& Salucci 2010).

\subsection{Galaxy sample and inferred properties}\label{sec|sample}

We now exploit the templates to derive global properties for a sample of $546$ local disk-dominated galaxies selected from the compilation by Persic \& Salucci (1995; see also Mathewson et al. 1992). For the selection we require the following properties: (i) the galaxy is Hubble-type classified as from $Sb$ to $Sm$ (to minimize the impact of bulge contamination to the stellar mass estimate), with an $I$-band magnitude $M_I$ in the range from $-18.5$ to $-23.8$ (for most of the sample $M_I$ falls in the range from $-19.5$ to $-22.5$); (ii) the galaxy has a well measured $I$-band photometry extended out to several disk scale-length $\ga 4\, R_d$, from which the optical radius $R_{\rm opt}\equiv 3.2\, R_d$ can be determined within $10\%$ accuracy; (iii) the rotation velocity $V_{\rm opt}$ at the optical radius is robustly measured at a few percent level. The ESO id. of the galaxies in the sample and the corresponding magnitude $M_I$, disk scale-length $R_d$, and optical velocity $V_{\rm opt}$ are reported in Table~\ref{table|data}. Note that we have updated the distance measurements for each galaxy of this sample by adopting the most recent determination from the NED database (see \texttt{https://ned.ipac.caltech.edu/}). As a further check on our  criteria aimed to select disk-dominated galaxies, we computed the $I-$band light concentration $C_I$ of the objects in the sample; this quantity is defined as the ratio of the radii containing $90\%$ and $50\%$ of the light. The related distribution is nearly Gaussian with peak $\langle C_I\rangle\approx 2.3$ and variance $\sigma_C\approx 0.1$. The average value is very close to the expectation $C_I\approx 2.32$ for a pure thin exponential disk; the range of $C_I$ spanned by the galaxies in our sample is also similar to that measured for disk-dominated galaxies in the SDSS (see Bernardi et al. 2010; their Fig. 4, referring to light concentration in the $r-$band).

For each galaxy in the sample, we compute the template with the values of the parameters $\kappa(M_I)$ and $\tilde R_c(M_I)$ appropriate for the galaxy magnitude $M_I$. The RC in physical units is recovered by de-normalizing the template $\tilde V_{\rm TOT}(\tilde R)$ with the observed values of $R_{\rm opt}$ and $V_{\rm opt}$ for each galaxy, to get the total $V_{\rm TOT}(R)$ and the associated disk $V_{\rm DISK}(R)$ and halo $V_{\rm HALO}(R)$ components. We have checked a posteriori that the denormalized template provides indeed a good description of the individual RCs; the distribution of reduced chi-squared residuals over the full sample has a mean around $1.1$ and a dispersion around $0.4$. Note that, on the other hand, our fitting procedure applied to the individual galaxy RCs (in place of the stacked data) gives, for most of the sample, very loose constraints on the parameters $\kappa$ and $\tilde R_c$ and hence on the stellar and halo masses.

From the disk component $V_{\rm DISK}(R)$, the stellar mass is straightforwardly derived as $M_\star\equiv V_{\rm DISK}^2(R_{\rm opt})\, R_{\rm opt}/1.1\, G$. In our analysis, we shall also use the central surface density in stars $\Sigma_0\equiv M_d/2\pi\, R_d^2$, and the effective radius $R_e\approx 1.68\, R_d\approx 0.52\, R_{\rm opt}$ wherein half of the stellar mass is encompassed. Finally, the specific angular momentum (per unit mass) of the stellar component is simply given by
\begin{equation}\label{eq|jstar}
j_\star\equiv 2\,f_R\, R_d\, V_{\rm opt}
\end{equation}
in terms of a shape factor $f_R\equiv \int{\rm d}x\, x^2\, e^{-x}\,V_{\rm TOT}(x R_d)/2\,V_{\rm opt}$ of order unity, that mildly depends on the shape of the RC template; for most of the galaxies in our sample, $f_R\simeq 0.93$ holds to an extremely good approximation.

As for the DM mass, we proceed as follows. From the dark halo component $V_{\rm HALO}(R)$, we derive the radius $R_{\Delta}$ where the average halo density $\langle\rho_{\Delta}\rangle\equiv 3\,M_{\Delta}(<R_{\Delta})/4\pi\, R_{\Delta}^3 = 3\, V_{\rm HALO}^2(R_\Delta)/4\pi\, G\, R_{\Delta}^2$ equals a given overdensity $\Delta$ of the critical one $\rho_{\rm crit}\equiv 3\,H^2/8 \pi\, G$, with $H$ the Hubble parameter. Setting $\langle\rho_{\Delta}\rangle=\Delta\, \rho_{\rm crit}$ yields the implicit equation $V_{\rm HALO}(R_\Delta)/R_\Delta=H\,\sqrt{\Delta/2}$,
that is easily solved for $R_\Delta$; the associated halo mass is simply estimated from $M_\Delta\equiv \Delta\, H^2\, R_\Delta^3/2\, G$.
For the standard cosmology adopted here, the quantities $R_{100}$ and $M_{100}$ corresponding to $\Delta=100$ constitute a very good approximation (e.g., Eke et al. 1996; Bryan \& Norman 1998) to the virial radius and mass of halos at redshift $z\approx 0$; hereafter we use the notations $R_{\rm H}=R_{100}$, $M_{\rm H}=M_{100}$ and $V_{\rm H}^2=G\, M_{\rm H}/R_{\rm H}$; in the literature, the radius
$R_{200}\approx 0.77\, R_{100}$, mass $M_{200}\approx 0.87\, M_{100}$ and velocity $V_{200}=(G\, M_{200}/R_{200})^{1/2}\approx 1.07\, V_{\rm H}$ are also often exploited. Finally, the specific angular momentum of the halo is defined as (see Mo et al. 1998, 2010)
\begin{eqnarray}\label{eq|jhalo}
\nonumber j_{\rm H} &=& \sqrt{2}\, \lambda\, R_{\rm H}\, V_{\rm H}\approx\\ 
\nonumber &\approx & 4.8\times 10^4\,\lambda\, E(z)^{-1/6}\, \left(M_{\rm H}\over 10^{12}\, M_\odot\right)^{2/3}~~{\rm km~s}^{-1}~{\rm kpc}\approx \\
& &\\
\nonumber& \approx& 2.9\times 10^{4}\, \lambda\, E(z)^{-1/2}\, \left(V_{\rm H}\over 100\, {\rm km~s}^{-1}\right)^{2}~~{\rm km~s}^{-1}~{\rm kpc}~,
\end{eqnarray}
where $E(z)=(H/H_0)^2\equiv \Omega_M\, (1+z)^3+\Omega_\Lambda$ is a redshift dependent factor (close to unity for local spirals), and $\lambda$ is the spin parameter of the host DM halo. Numerical simulations (see Barnes \& Efstathiou 1987; Bullock et al. 2001; Macci\'o et al. 2007; Zjupa \& Springel 2017) have shown that $\lambda$ exhibits a log-normal distribution with average value $\langle\lambda\rangle\approx 0.035$ and dispersion $\sigma_{\log \lambda} \approx 0.25$ dex, nearly independent of mass and redshift.

\section{Results}\label{sec|results}

In this Section we discuss the relationships between global quantities of the halo and the stellar components for our sample, and compare them with the outcomes of independent datasets and with the expectation from the classic theory of disk formation.

\subsection{Dark and stellar masses}\label{sec|Mh_Mstar}

\begin{figure*}
\epsscale{1}\plotone{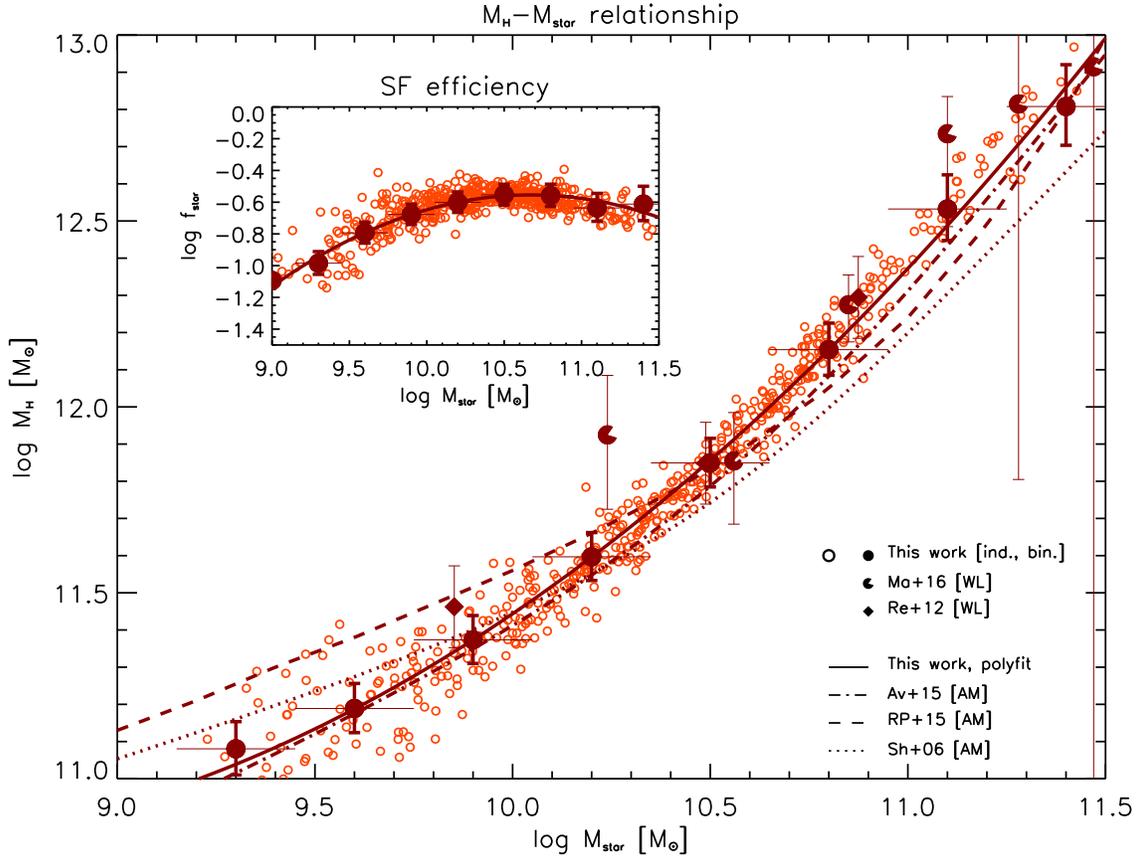}\caption{Relationship between the halo and stellar mass derived from our analysis, for individual galaxies (small circles) and for the mean of binned data (big circles); for the latter the thin horizontal errorbars show the binsize in $\log M_\star$ and the thick vertical errorbars show the corresponding $1\sigma$ dispersion around the mean. Solid line is a polynomial fit to the mean for binned data from this work. Weak lensing determinations by Mandelbaum et al. (2016; pacmans) and by Reyes et al. (2012; diamonds), and (sub-)halo abundance matching outcomes by Aversa et al. (2015; dot-dashed line), by Rodriguez-Puebla et al. (2015; dashed line) and by Shankar et al. (2006; dotted line) are also reported. The inset represents the corresponding relation between the star formation efficiency $f_\star\equiv M_\star/f_b\,M_{\rm H}$ and the stellar mass $M_\star$.}\label{fig|Mh_Mstar}
\end{figure*}

In Fig.~\ref{fig|Mh_Mstar} we show the relationship between the halo mass $M_{\rm H}$ and the stellar mass $M_\star$ from our analysis, both for the individual galaxies of the sample, and for the mean within bins of given $M_\star$. We also show a polynomial fit to our mean result for binned data, whose parameters are reported in Table~\ref{table|fits}.

We compare the relationship from our analysis to the independent determinations via weak gravitational lensing by Mandelbaum et al. (2016) and Reyes et al. (2012), and via abundance matching techniques (in terms of the average $\langle M_{\rm H}\rangle$ at given $M_\star$) by Shankar et al. (2006), Aversa et al. (2015), Rodriguez-Puebla et al. (2015; see their Fig.~10), finding a remarkable agreement. Note that we derive dynamical estimates of the stellar mass for a well-defined sample of pure disks, while the above weak lensing studies refer to stellar masses inferred from photometry of blue color-selected galaxies. We stress that the samples and methodologies are completely different (and suffer of different biases), therefore the apparent agreement is of paramount relevance in order to check for the respective systematics and assess the relationship between stellar mass and halo mass in disk-dominated galaxies. In particular, our basic assumptions on the halo component (see Sect.~\ref{sec|URC}) are validated and complemented by weak lensing measurements, which are mostly sensitive to the DM distribution in the outskirts.

In addition, it is striking that the $1\sigma$ dispersion around the mean relation from our analysis amounts to about $0.08$ dex. This is appreciably smaller than in any other previous determination, and sets a new standard for the $M_{\rm H}-M_\star$ relationship of local spiral galaxies. Hereafter the quoted dispersions around the mean are meant to include the intrinsic scatter of the data for individual galaxies within the bin, the measurements errors in the determination of the optical radii $R_{\rm opt}$ and velocities $V_{\rm opt}$ and, as major contribution, the uncertainty in the reconstruction of the templates (associated to the uncertainties in the parameters $\kappa$ and $\tilde R_c$, cf. Fig.~\ref{fig|URC_parfit}).

The inset illustrates the same outcome in terms of the star formation efficiency $f_\star\equiv M_{\star}/f_b\, M_{\rm H}\approx M_{\star}/0.16\, M_{\rm H}$, i.e., the efficiency at which the original baryon content of a halo is converted into stars. This is often considered a fundamental quantity, since it bears the imprint of the processes (gas cooling and condensation, star formation, energy/momentum feedback from supernovae and stellar winds, etc.) ruling galaxy formation and evolution. We find that it amounts to around $30\%$ for stellar masses $M_\star\ga$ a few $10^{10}\, M_\odot$, which is quite an high value if compared to the maximum of $\la 20\%$ for early-type galaxies (see Shi et al. 2017 and reference therein).

The physical interpretation is that the stellar masses of late-type galaxies are being accumulated at low rates over several Gyrs, regulated by continuous energy feedback from supernova explosions and stellar winds; this is to be contrasted with the much larger star formation rates and short duration timescales $\la$ Gyr (likely related to feedback from the central accreting supermassive black hole) for the progenitor of massive ellipticals (e.g., Lapi et al. 2011, 2017; Glazebrook et al. 2017; Kriek et al. 2017).
Note that the low metallicity of late-type galaxies relative to ellipticals calls for processes able to continuously dilute the star-forming gas; viable possibilities are galactic fountains and/or continuous inflow of pristine gas from the outer regions of the dark halo (see Fraternali 2017; Shi et al. 2017).

Although our analysis here concerns only local spirals, Hudson et al. (2015) showed that the relationship between stellar and halo masses does not appreciably evolve out to $z\approx 0.7$. This is consistent with the aforementioned notion that, in late-type galaxies, the star formation and DM accretion goes in parallel along cosmic times. It would be extremely important to further test this finding by independent measurements, and in particular by applying our analysis to a statistically significant high-redshift sample of disk-dominated galaxies. If the shape and tightness of the $M_{\rm H}$ vs. $M_\star$ relation stay put or evolve in a controlled way with respect to the local Universe, then the halo masses of high-redshift spiral galaxies out to $z\sim 1$ could be robustly determined by evaluating the stellar mass content, so dispensing with the systematic uncertainties related to clustering or to weak gravitational lensing measurements. In perspective, a reliable determination of halo masses in the redshift range $z\la 1$, an epoch where the dark energy starts driving the accelerated expansion of the Universe, could potentially have profound implications for cosmological studies.

\subsection{Stellar size and surface density}\label{sec|sizes}

\begin{figure*}
\epsscale{1}\plotone{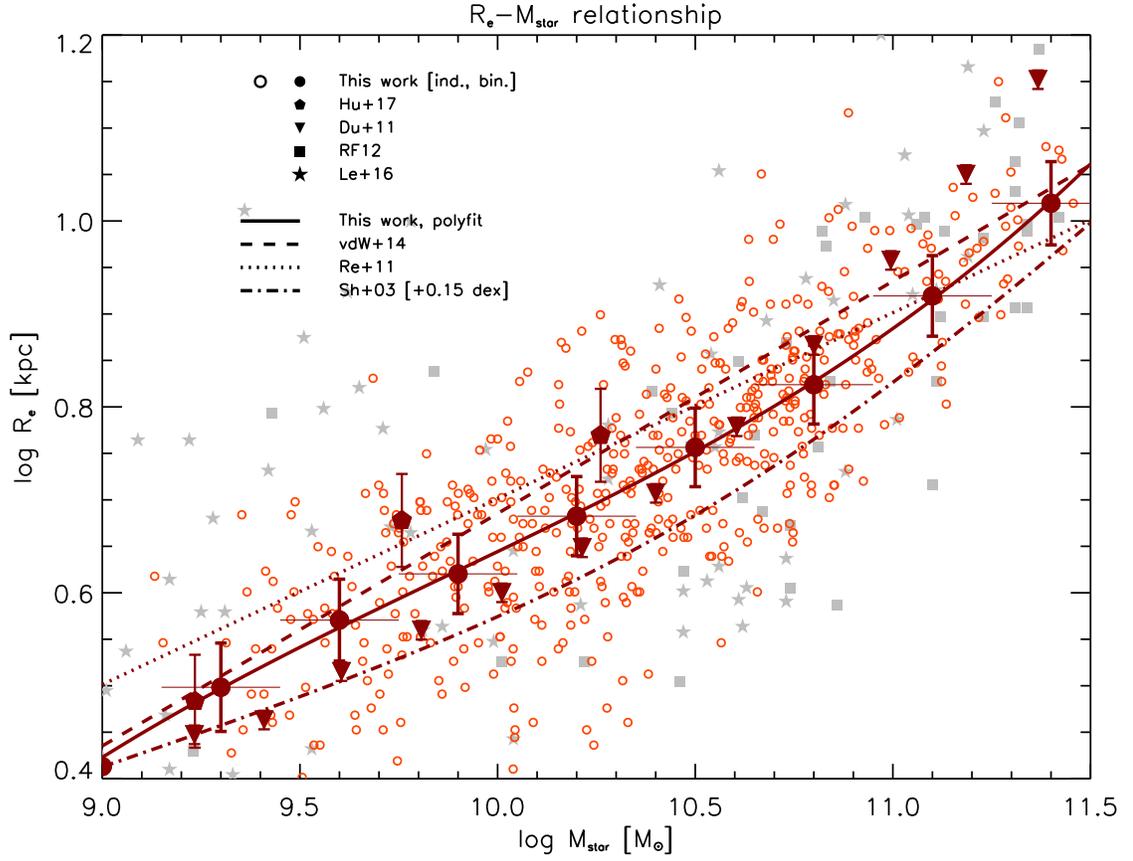}\caption{Relationship between the stellar effective half-mass radius $R_e$ and the stellar mass derived from our analysis, for individual galaxies (small circles) and for the mean of binned data (big circles); for the latter the thin horizontal errorbars show the binsize in $\log M_\star$ and the thick vertical errorbars show the corresponding $1\sigma$ dispersion around the mean. Data from individual RC modeling by Romanowsky \& Fall (2012; squares) and by Lelli et al. (2016; stars) are reported. Determinations from photometric data and SED modeling by Huang et al. (2017; pentagons) and by Dutton et al. (2011; inverse triangles) are also shown. Solid line is a polynomial fit to the mean for binned data from this work, dashed line is the observational determination by van der Wel et al. (2014), dotted line is by Reyes et al. (2011), and dot-dashed line is by Shen et al. (2003;  their $\log R_e$ have been corrected by $+0.15$ dex to transform from circularized to major-axis sizes).}\label{fig|Re_Mstar}
\end{figure*}

In Fig.~\ref{fig|Re_Mstar} we show the relationship between the effective radius $R_e\approx 1.68\, R_d$ of the stellar component and the stellar mass $M_\star$ from our analysis, both for the individual galaxies of the sample and for the mean within bins of given $M_\star$. We also show a polynomial fit to our mean results for binned data, whose parameters are reported in Table~\ref{table|fits}; this is in good agreement with the relationships for late-type galaxies by Reyes et al. (2011), by van der Wel et al. (2014) and by Shen et al. (2003; their $\log R_e$ have been corrected by $+0.15$ dex to transform from circularized to major-axis sizes).

We compare the relationship from our analysis to the direct determinations via photometry and SED modeling by Dutton et al. (2011) and Huang et al. (2017), finding a remarkable agreement. We also confront our relation with the independent measurements from individual RC modeling by Romanowsky \& Fall (2012) and Lelli et al. (2016), finding a good accord. The $1\sigma$ dispersion around the mean relation from our analysis amounts to about $0.05$ dex. Note that the small scatter in our $R_e$ vs. $M_\star$ relationship is partly due to the selection of pure disk-dominated galaxies in the sample analyzed here.

\begin{figure*}
\epsscale{1}\plotone{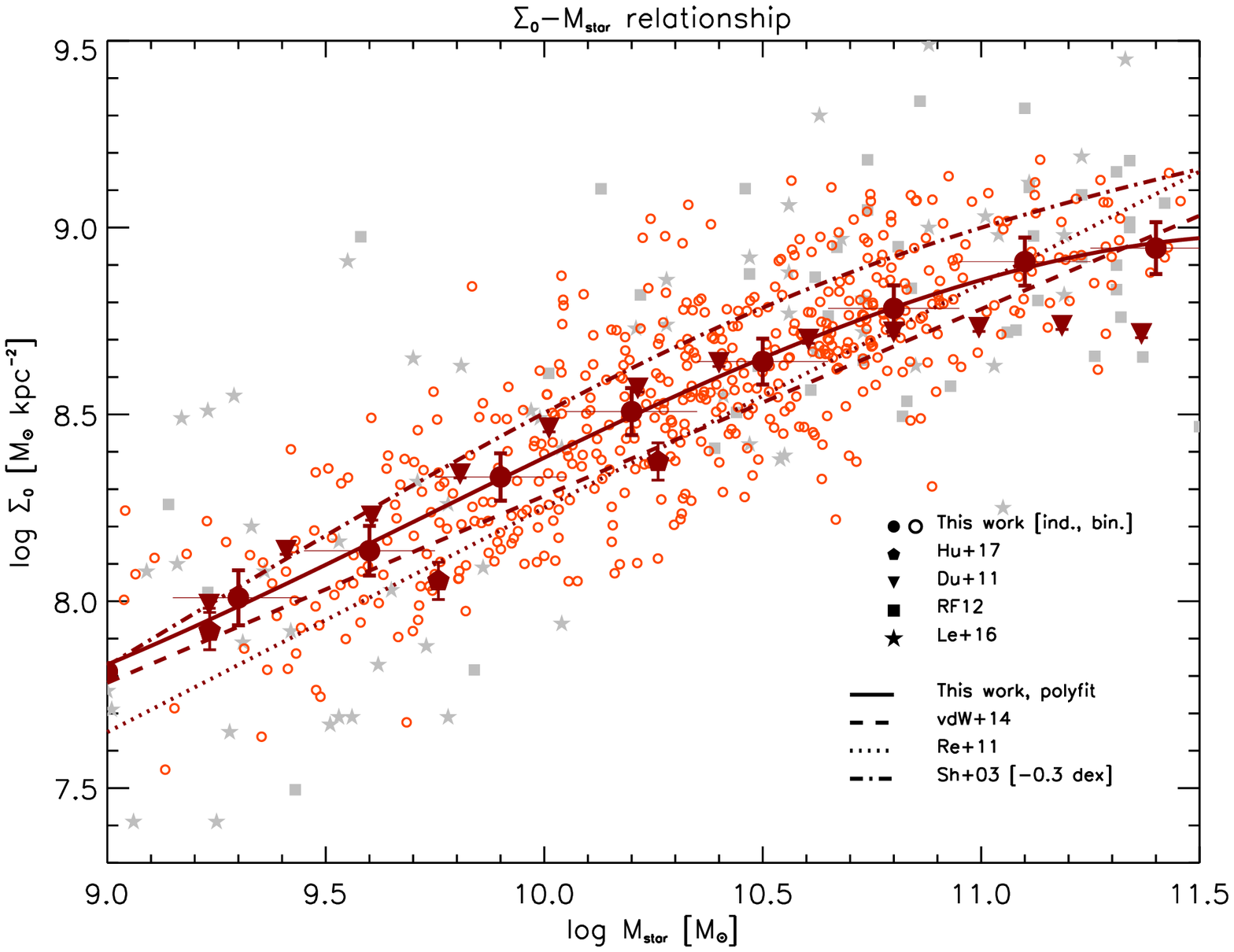}\caption{Relationship between the central surface brightness $\Sigma_0$ and the stellar mass $M_\star$ derived from our analysis, for individual galaxies (small circles) and for the mean of binned data (big circles); for the latter the thin horizontal errorbars show the binsize in $\log M_\star$ and the thick vertical errorbars show the corresponding $1\sigma$ dispersion around the mean. Data from individual RC modeling by Romanowsky \& Fall (2012; squares) and by Lelli et al. (2016; stars) are reported. Determinations by photometry and SED modeling by Huang et al. (2017; pentagons) and by Dutton et al. (2011; inverse trinagles) are also illustrated. Solid line is a polynomial fit to the mean for binned data from this work, dashed line is the relation observed by van der Wel et al. (2014), dotted line is by Reyes et al. (2011), and dot-dashed line is by Shen et al. (2003; their $\log \Sigma_0\sim -2\,\log R_e $ have been corrected by $-0.3$ dex to transform from circularized to major-axis sizes).}\label{fig|Sigma0_Mstar}
\end{figure*}

The same data can be recast in terms of the relationship between the central surface density $\Sigma_0\equiv M_\star/2\pi\,R_d^2$ and the stellar mass $M_\star$, which is plotted in Fig.~\ref{fig|Sigma0_Mstar}. We stress that the mean relationship flattens toward values $\Sigma_0\sim 10^9\, M_\odot$ kpc$^2$ for high stellar masses $M_\star\ga$ a few $10^{10}\, M_\odot$. The physical interpretation is that to reach higher stellar surface densities, large masses must accumulate (rapidly) within quite small sizes, a condition regularly met in the progenitors of compact spheroids (see Lapi et al. 2011, 2017; Glazebrook et al. 2017; Kriek et al. 2017) but difficult to achieve in extended disks with quiet star formation histories.

\subsection{Tully-Fisher relation and $M_\star/L$ ratios}\label{sec|MoverL}

\begin{figure*}
\epsscale{1}\plotone{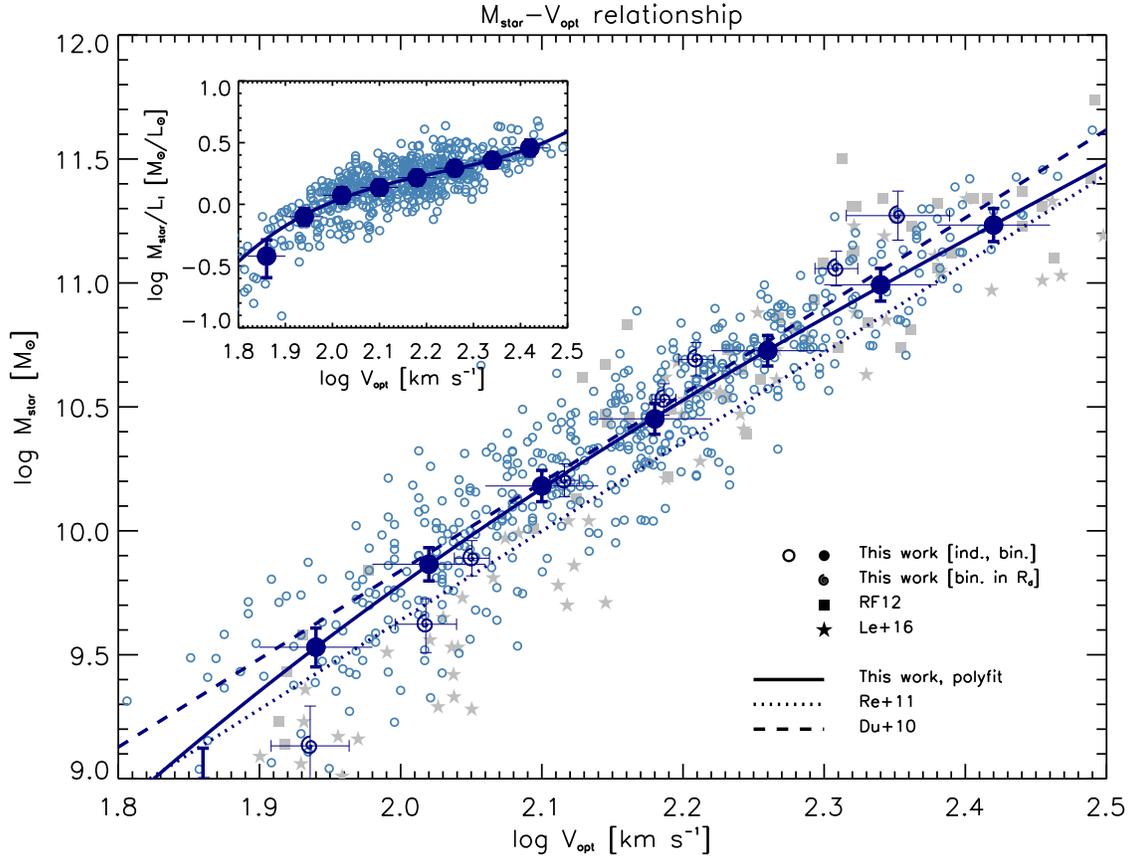}\caption{Relationship between the stellar mass $M_\star$ and the optical velocity $V_{\rm opt}$ derived from our analysis, for individual galaxies (small circles) and for the mean of binned data (big circles); for the latter the thin horizontal errorbars show the binsize in $\log V_{\rm opt}$ and the thick vertical errorbars show the corresponding $1\sigma$ dispersion around the mean. The relationship for data binned in $\log R_d$ is also shown (spirals). Data from individual RC modeling by Romanowsky \& Fall (2012; squares) and by Lelli et al. (2016; stars) are reported. Solid line is a polynomial fit to the mean for binned data from this work, dotted line is the relation observed by Reyes et al. (2011), and dashed line is by Dutton et al. (2010) for local spiral galaxies. The inset shows the corresponding mass-to-light ratios $M/L_I$ from our  analysis.}\label{fig|Mstar_Vopt}
\end{figure*}

In Fig.~\ref{fig|Mstar_Vopt} we show the Tully-Fisher relationship between the stellar mass $M_\star$ and the optical velocity $V_{\rm opt}$. We plot the outcome from our analysis, both for the individual galaxies of the sample and for the mean within bins of given $V_{\rm opt}$.
We also show a polynomial fit to our mean results for binned data, whose parameters are reported in Table~\ref{table|fits}. This is in good agreement, especially for $V_{\rm opt}\ga 100$ km s$^{-1}$, with the classic powerlaw determinations $M_\star\propto V_{\rm opt}^{3.7}$ by Dutton et al. (2010) and Reyes et al. (2011). The $1\sigma$ dispersion around the mean relation from our analysis amounts to about $0.08$ dex, mainly contributed by the uncertainty in template reconstruction. We also confront the outcome of our analysis with independent results from individual RC modeling by Romanowsky \& Fall (2012) and Lelli et al. (2016), finding a pleasing consistency.

We stress that the latter authors adopt a definite value of $M_\star/L$ in their analysis; specifically Romanowsky \& Fall (2012) assume $M_\star/L_K\approx 1$ while Lelli et al. (2016) assume $M_\star/L_{3.6\, \mu{\rm m}}\approx 0.5$. On the other hand, the ratio $M_\star/L_I$ constitutes an outcome of our approach; the resulting values are plotted in the inset of Fig.~\ref{fig|Mstar_Vopt} as a function of $V_{\rm opt}$, and range from $M_\star/L_I\approx 1$ for $V_{\rm opt}\sim 100$ km s$^{-1}$ to $M_\star/L_I\approx 3$ for $V_{\rm opt}\ga 200$ km s$^{-1}$, and to $M_\star/L_I\la 0.5$ for $V_{\rm opt}\la 70$ km s$^{-1}$. Note that the $1\sigma$ dispersion of the $M_\star/L_I$ values for individual galaxies and the dispersion around the mean are consistent with those of the $M_\star-V_{\rm opt}$ relationship. Despite the difficulty in soundly comparing the $M_\star/L$ values computed by us to those adopted by Romanowsky \& Fall (2012) and Lelli et al. (2016), it is worth stressing that the corresponding stellar masses are consistent. However, the assumption of a $M_\star/L$ ratio constant with $V_{\rm opt}$ is likely at the origin of the slightly larger scatter in their stellar masses (mirrored in all relationships involving $M_\star$) with respect to those resulting from our analysis. In addition, we notice that our dynamical determinations of $M/L_I$ ratios are quite close to that adopted by Bell et al. (2003) and Reyes et al. (2011) based on $(g-r)$ colors for disk galaxies, converted to a Chabrier IMF.

We also check the dependence of the derived Tully-Fisher relationship on the disk scale-length (or equivalently on the effective radius). To this purpose, in Fig.~\ref{fig|Mstar_Vopt} we illustrate the average relationship when grouping the data in bins of $\log R_d$; the outcome is, within uncertainties, indistinguishable from the overall $M_\star$ vs. $V_{\rm opt}$ relation. Such an independence of the size (and consequently of the disk surface brightness) has been originally pointed out by Courteau et al. (2007).

\begin{figure*}
\epsscale{1}\plotone{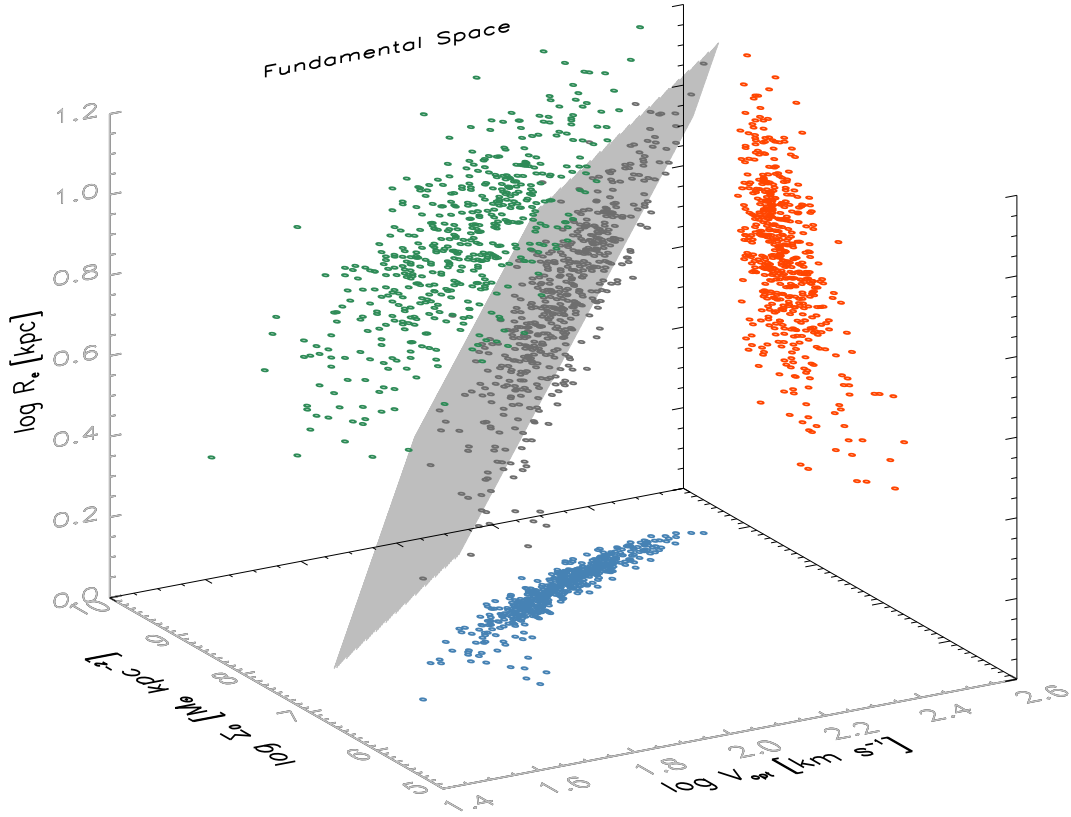}\caption{Fundamental space of spiral galaxies, involving the effective half-mass radius $R_e$, central stellar surface density $\Sigma_0$ and the optical velocity $V_{\rm opt}$ derived from our analysis for individual galaxies (small circles) and the projected relationships in the $R_e-\Sigma_0$ (red), $R_e-V_{\rm opt}$ (green), and $\Sigma_0-V_{\rm opt}$ (blue) planes. The shaded surface illustrates the best-fit plane to the data for individual galaxies from a principal component analysis.}\label{fig|FP}
\end{figure*}

In Fig.~\ref{fig|FP} we present the fundamental space of spiral galaxies, involving three observable quantities: the effective radius $R_e$, the central stellar surface density $\Sigma_0$ and the optical velocity $V_{\rm opt}$, as derived from our analysis; we also show the projected relationships in the $R_e-\Sigma_0$, $R_e-V_{\rm opt}$, and $\Sigma_0-V_{\rm opt}$ planes. The shaded surface illustrates the best-fit plane to the datapoints for individual galaxies from a principal component analysis, which is given by the expression
\begin{eqnarray}
\nonumber \log R_e [{\rm kpc}] &\approx& 1.31-0.57\,\log \Sigma_0 [M_\odot~{\rm kpc}^{-2}]+\\
\\
\nonumber &+&1.99\,\log V_{\rm opt} [{\rm km~s}^{-1}]~;
\end{eqnarray}
this is analogous to the fundamental plane of ellipticals (see Djorgovski \& Davis 1987; Dressler et al. 1987). At variance with the latter, the plane of spirals is substantially tilted with respect to all the axes, and so its projections are characterized by an appreciable dispersion; moreover, it does not provide a comparably good rendition of the data distribution in the full three dimensional space.

\subsection{Optical to virial velocity ratio}\label{sec|velratio}

\begin{figure*}
\epsscale{1}\plotone{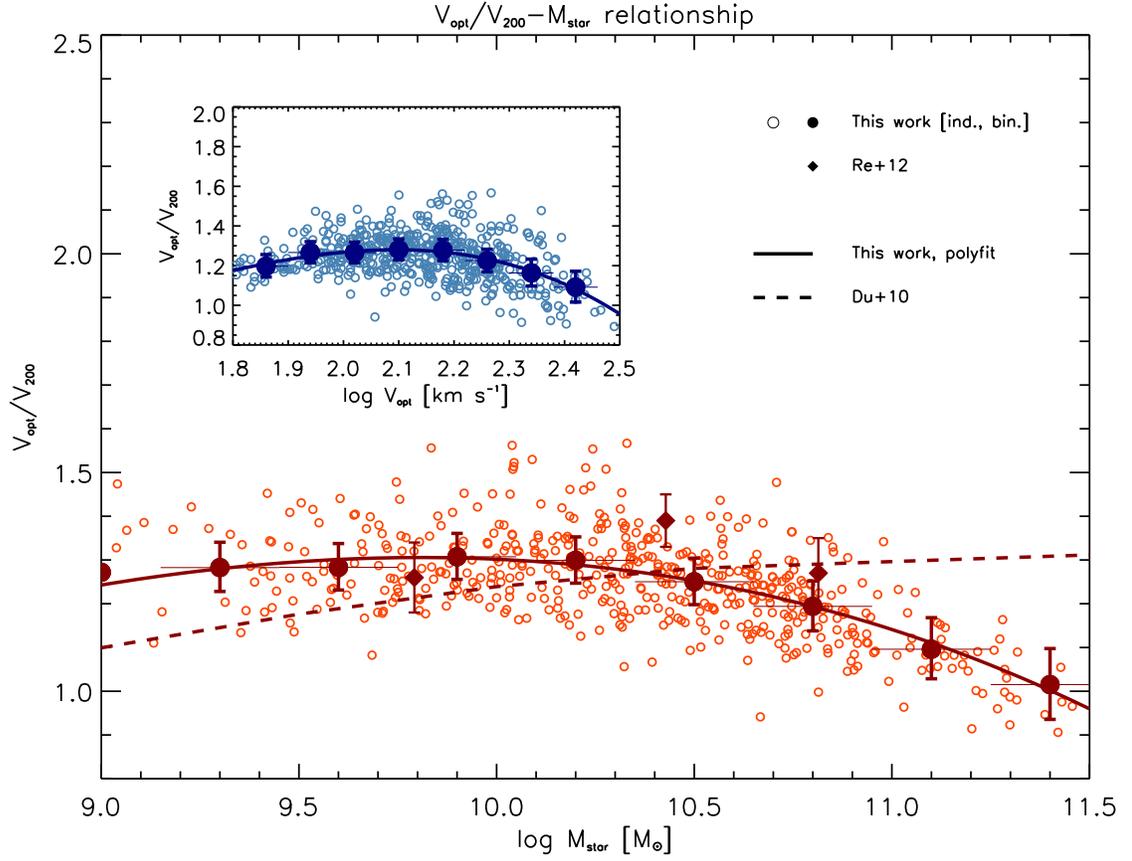}\caption{Relationship between the ratio $V_{\rm opt}/V_{200}$ and the stellar mass $M_\star$ derived from our analysis, for individual galaxies (small circles) and for the mean of binned data (big circles); for the latter the thin horizontal errorbars show the binsize in $\log M_\star$ and the thick vertical errorbars show the corresponding $1\sigma$ dispersion around the mean. Determinations by Reyes et al. (2012; diamonds) via abundance matching are also reported. Solid line is a polynomial fit to the mean for binned data from this work, and dashed line is the relation by Dutton et al. (2010). The inset shows the corresponding relation between the ratio $V_{\rm opt}/V_{200}$ and the optical velocity $V_{\rm opt}$.}\label{fig|Vratio_Mstar}
\end{figure*}

In Fig.~\ref{fig|Vratio_Mstar} we show the relationship between the stellar mass $M_\star$ and the ratio $V_{\rm opt}/V_{200}$ of the optical velocity to the halo circular velocity at the radius where the overdensity is $200$ times the critical one. We plot the outcome from our analysis, both for the individual galaxies of the sample and for the mean within bins of given $M_\star$. We also show a polynomial fit to our mean results for binned data, whose parameters are reported in Table~\ref{table|fits}; this is consistent with the relationship by Dutton et al. (2010) for late-type galaxies. We also compare our relationship to the independent determinations via weak galaxy-galaxy lensing by Reyes et al. (2012), finding a remarkable agreement. The $1\sigma$ dispersion around the mean relation from our analysis amounts to about $0.06$ dex, making our determination one of the most precise to date.

The weak dependence of the ratio $V_{\rm opt}/V_{200}$ on $M_\star$
implies that the optical velocity $V_{\rm opt}$ is a good proxy of the halo circular velocity $V_{200}$ in the outskirts. The inset of Fig.~\ref{fig|Vratio_Mstar} shows the same velocity ratio $V_{\rm opt}/V_{200}$ as a function of the optical velocity $V_{\rm opt}$; such a relation could be used to infer halo masses with high accuracy. In Sect.~\ref{sec|RdVopt} we will exploit this results to theoretically interpret the $R_d-V_{\rm opt}$ relationship.

\subsection{Specific angular momentum}\label{sec|jstar}

\begin{figure*}
\epsscale{1}\plotone{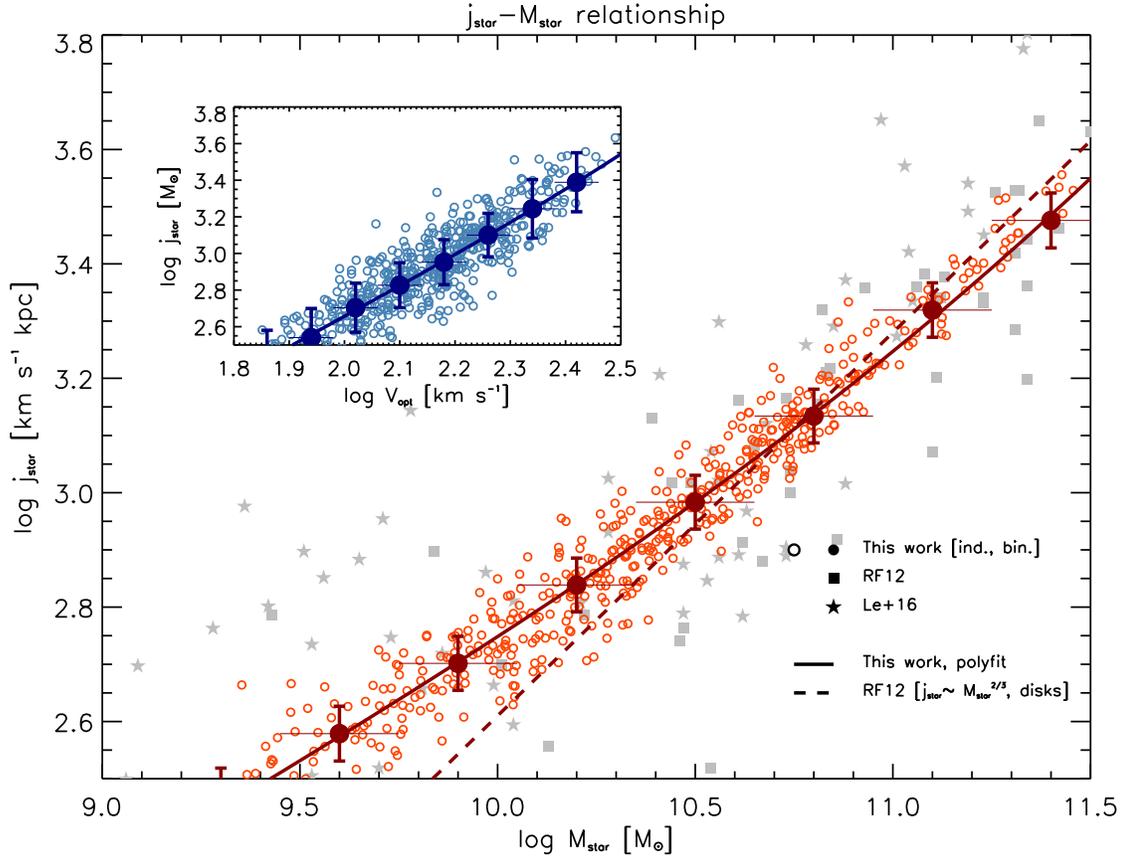}\caption{Relationship between the stellar specific angular momentum $j_\star$ and the stellar mass $M_\star$ derived from our analysis, for individual galaxies (small circles) and for the mean of binned data (big circles); for the latter the thin horizontal errorbars show the binsize in $\log M_\star$ and the thick vertical errorbars show the corresponding $1\sigma$ dispersion around the mean. Data from individual RC modeling by Romanowsky \& Fall (2012; squares) and by Lelli et al. (2016; stars) are reported. Solid line is a polynomial fit to the mean for binned data from this work, dashed line is the relation with fixed slope $j_\star\propto M_\star^{2/3}$ for pure disks by Romanowsky \& Fall (2012). The inset shows the corresponding relation between $j_\star$ and $V_{\rm opt}$ from our analysis.}\label{fig|jstar_Mstar}
\end{figure*}

In Fig.~\ref{fig|jstar_Mstar} we illustrate the relationship between the specific angular momentum $j_\star$ of the stellar component (see Eq.~[\ref{eq|jstar}]) and the stellar mass $M_\star$. We plot the outcome from our analysis, both for the individual galaxies of the sample and for the mean within bins of given $M_\star$. We also show a polynomial fit to our mean result for binned data, whose parameters are reported in Table~\ref{table|fits}. The dashed line is the relation with fixed slope $j_\star\propto M_\star^{2/3}$ for pure disks by Romanowsky \& Fall (2012). It is seen that our relation features such a $2/3$ slope at high masses but then flattens out for $M_{\rm H}\la$ a few $10^{10}\, M_\odot$; this is a consequence of the trivial scaling $j_\star\propto j_{\rm H}\propto M_{\rm H}^{2/3}$ from Eq.~\ref{eq|jhalo} and angular momentum conservation, coupled with the behavior of the $M_{\rm H}-M_\star$ relation (see Fig.~\ref{fig|Mh_Mstar}) that is log-linear at high stellar masses and flattens appreciably toward smaller ones.

We confront our result with the independent measurements from individual RC modeling of disk-dominated galaxies by Romanowsky \& Fall (2012) and Lelli et al. (2016), finding good agreement, though the scatter of their datapoints is appreciably larger especially for massive spirals (likely due to their assumption of a $M_\star/L$ ratio constant with luminosity). In fact, the $1\sigma$ dispersion around the mean $j_\star-M_\star$ relation from our analysis amounts to about $0.05$ dex.  The inset of Fig.~\ref{fig|jstar_Mstar} shows the corresponding relation between $j_\star$ and the optical velocity $V_{\rm opt}$, that will be used in Sect.~\ref{sec|RdVopt} to theoretically interpret the $R_d-V_{\rm opt}$ relationship.

\begin{figure*}
\epsscale{1}\plotone{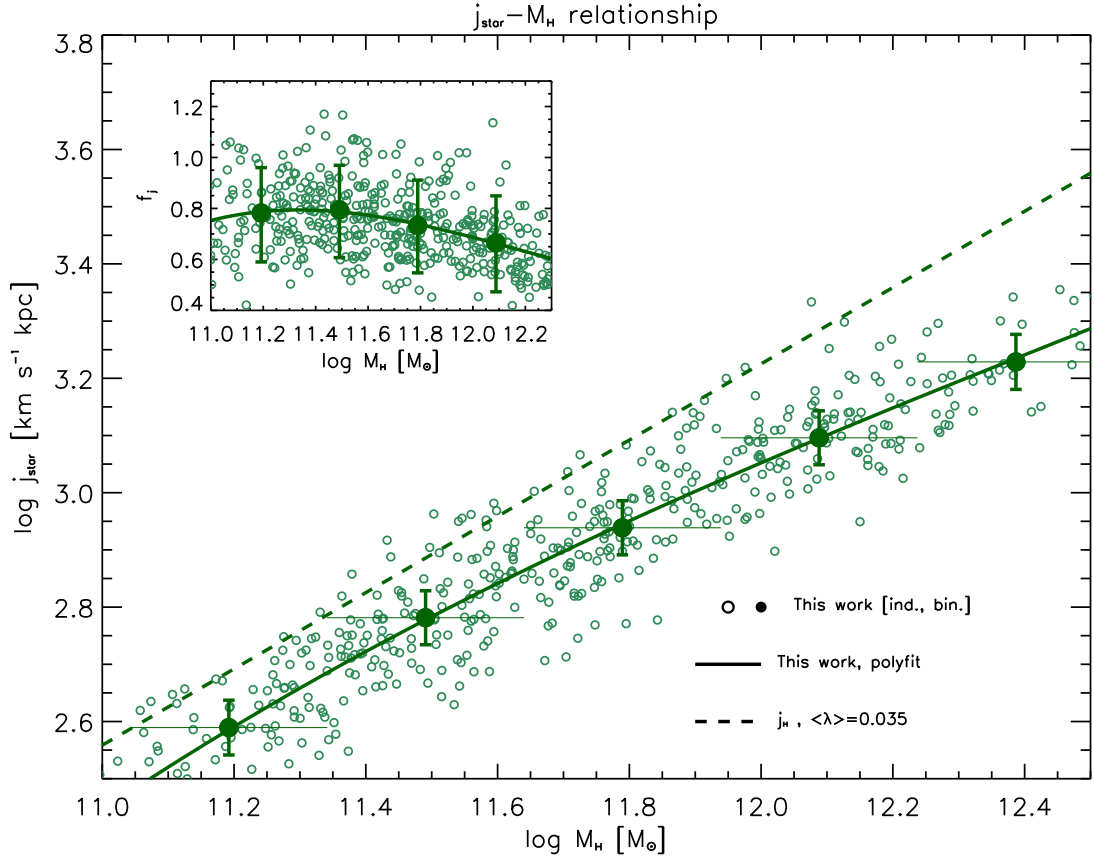}\caption{Relationship between the stellar specific angular momentum $j_\star$ and the halo mass $M_{\rm H}$ derived from our analysis, for individual galaxies (small circles) and for the mean of binned data (big circles); for the latter the thin horizontal errorbars show the binsize in $\log M_{\rm H}$ and the thick vertical errorbars show the corresponding $1\sigma$ dispersion around the mean. Solid line is a polynomial fit to the mean for binned data from this work. The inset shows the corresponding retention/sampling factor $f_j\equiv j_\star/j_{\rm H}$ as a function of the halo mass $M_{\rm H}$.}\label{fig|jstar_Mh}
\end{figure*}

Fig.~\ref{fig|jstar_Mh} present $j_\star$ as a function of the halo mass $M_{\rm H}$. The $1\sigma$ dispersion around the mean relation from our analysis amounts to about $0.05$ dex. The tightness of the $j_\star-M_{\rm H}$ relation suggests the possibility of inferring halo masses via the specific angular momentum $j_\star\sim 2\, V_{\rm opt}\, R_d$, which in turn can be accurately estimated by measurements of the optical velocity and of the disk scale-length (see also Romanowsky \& Fall 2012). The inset shows the fraction $f_j\equiv j_\star/j_{\rm H}$ representing the amount of the halo angular momentum (see Eq.~\ref{eq|jhalo}) still retained or sampled by the stellar component (see Shi et al. 2017). The standard and simplest theory of disk formation envisages sharing and conservation of the specific angular momentum between baryons (in particular, stars) and DM, to imply $f_j\approx 1$. Remarkably, our analysis shows that such a picture is closely supported by the data, since the values of $f_j$ range from $0.6$ to $1$, with a weak dependence on $M_{\rm H}$ (see also Romanowsky \& Fall 2012). Note that, from a theoretical perspective, a value $f_j$ below $1$ can be ascribed to inhibited collapse of the high angular momentum gas located in the outermost regions by stellar feedback processes (e.g., Fall 1983; Shi et al. 2017).

\subsection{Stellar to virial size ratio}\label{sec|ReR200c}

\begin{figure*}
\epsscale{1}\plotone{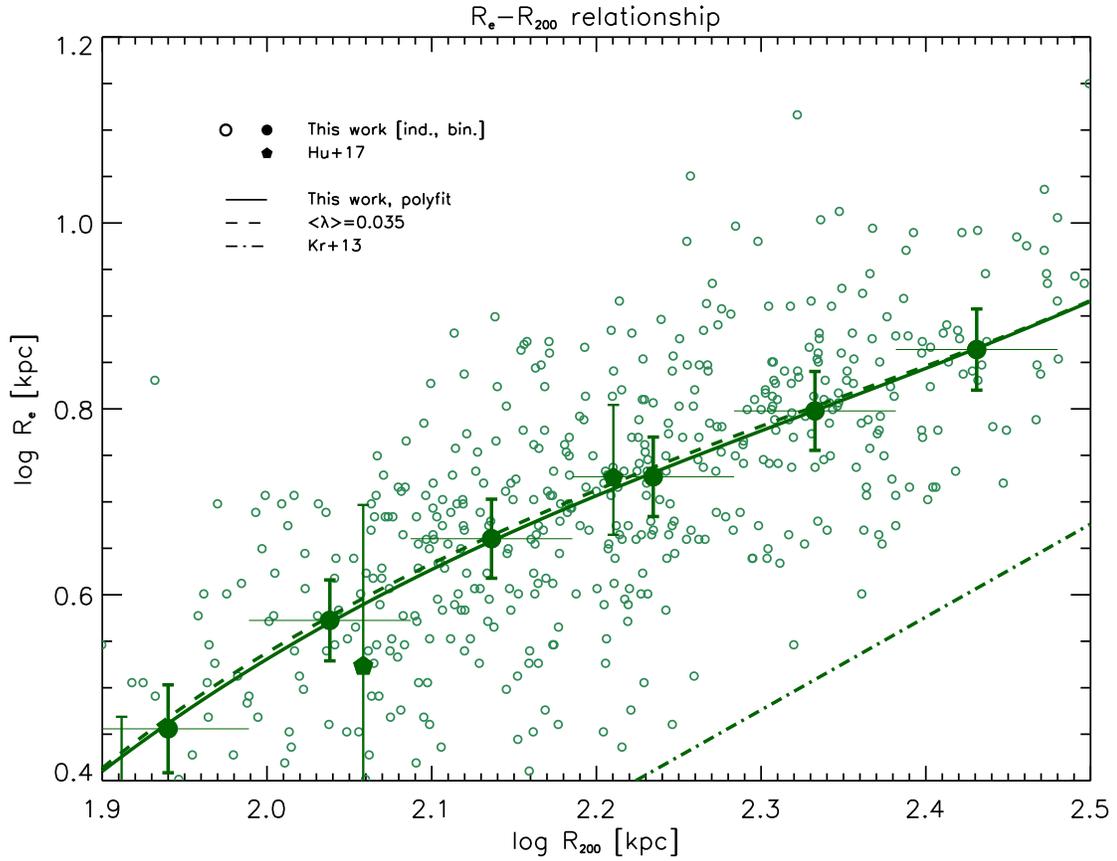}\caption{Relationship between the stellar effective radius $R_e$ and the halo size $R_{200}$ derived from our analysis, for individual galaxies (small circles) and for the mean of binned data (big circles); for the latter the thin horizontal errorbars show the binsize in $\log R_{200}$ and the thick vertical errorbars show the corresponding $1\sigma$ dispersion around the mean. Determinations by Huang et al. (2017; pentagons) via abundance matching are also shown. Solid line is a polynomial fit to the mean for binned data from this work, dashed line is the theoretical expectation based on angular momentum conservation (see Eq.~[\ref{eq|RdR200c}]), and dot-dashed line is the relation proposed by Kravtsov et al. (2013).}\label{fig|Re_R200c}
\end{figure*}

Fig.~\ref{fig|Re_R200c} shows the relationship between the stellar effective radius $R_e$ and the halo radius $R_{200}$ where the overdensity is $200$ times the critical value (see Sect.~\ref{sec|sample}). We illustrate the outcome from our analysis, both for the individual galaxies of the sample and for the mean within bins of given $R_{200}$. The polynomial fit to our mean results for binned data is also reported (see Table~\ref{table|fits}). Our mean relation is in remarkable agreement with the determination via abundance matching by Huang et al. (2017); on the other hand, it is substantially higher than the relation proposed by Kravtsov (2013), whose sample is actually dominated by early-type galaxies.

It is also interesting to compare our relation to that expected from the classic theory of disk formation, envisaging sharing and conservation of specific angular momentum between DM and stars (see Mo et al. 1998, 2010). To this purpose, we equalize $j_\star=f_j\, j_{\rm H}$ with $j_{\rm H}\equiv \sqrt{2}\, \lambda\, R_{\rm H}\, V_{\rm H}$ from Eq.~(\ref{eq|jhalo}) to $j_\star=2\, f_R\, R_d\, V_{\rm opt}$ from Eq.~(\ref{eq|jstar}). Introducing
$f_V\equiv V_{\rm opt}/V_{\rm H}$, we obtain the relation
\begin{equation}\label{eq|RdR200c}
R_d\simeq {\lambda\over \sqrt{2}}\,{f_j\over f_R\,f_V}\, R_{\rm H}~,
\end{equation}
where fits to the dependencies of $f_j$ and $f_V\approx 1.07\,V_{\rm opt}/V_{200}$ on $R_{200}\approx 0.77\, R_{\rm H}$ are given in Table~\ref{table|fits}; note that the factor $f_j/f_R\,f_V$, often neglected in the literature, bends downward the relation for $R_{200}\ga 150$ kpc. For an average value of the spin parameter $\langle\lambda\rangle\approx 0.035$ as indicated by numerical simulations, the expectation of Eq.~(\ref{eq|RdR200c}) is in excellent agreement with our mean determination from data. We stress that the $1\sigma$ dispersion around the mean relation from our analysis amounts to about $0.05$ dex.

\subsection{Disk scale-length vs. optical velocity}\label{sec|RdVopt}

\begin{figure*}
\epsscale{1}\plotone{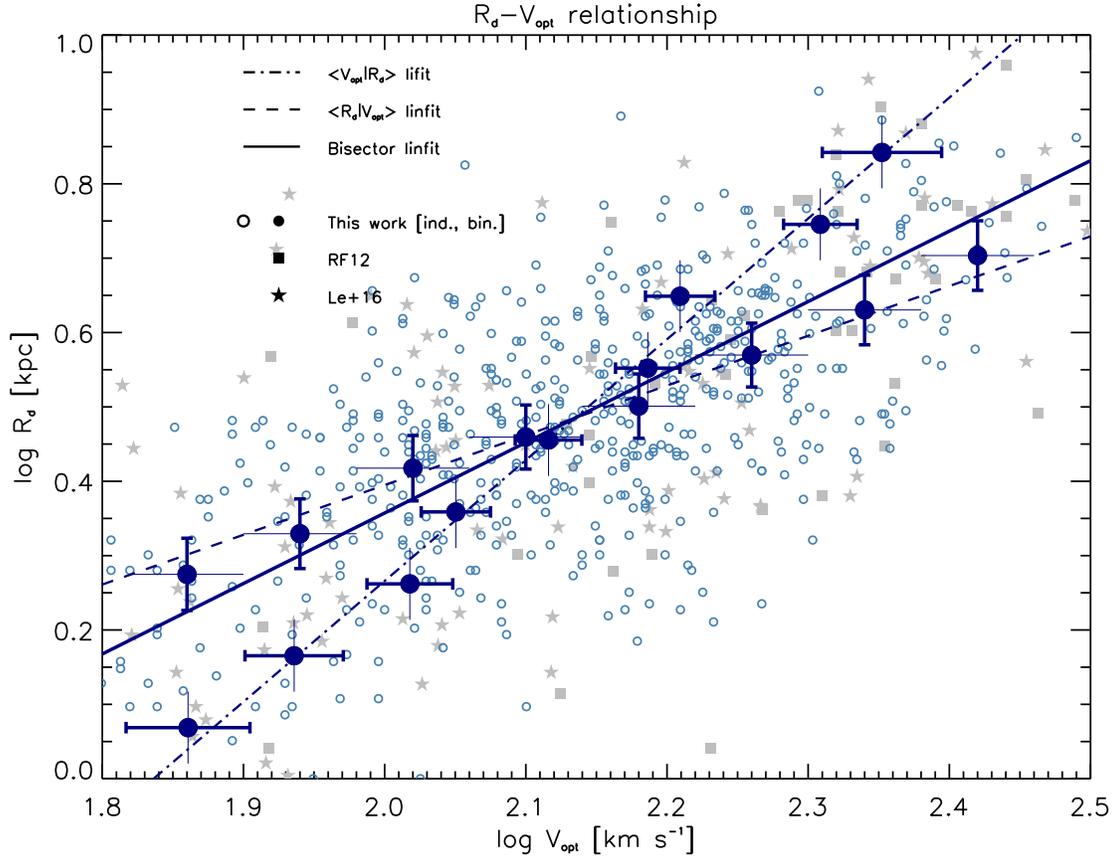}\caption{Relationships between the disk scale-length $R_d$ and the optical velocity $V_{\rm opt}$ derived from our analysis, for individual galaxies (small circles) and for the mean of binned data (big circles); for the latter the thin horizontal errorbars show the binsize in $\log V_{\rm opt}$ (or $\log R_d$) and the thick errorbars show the corresponding $1\sigma$ dispersion around the mean. Data by Romanowsky \& Fall (2012; squares) and by Lelli et al. (2016; stars) are reported. Dashed line is a linear fit to the mean $R_d-V_{\rm opt}$ relation and dot-dashed line to the mean $V_{\rm opt}-R_d$ relation for binned data from this work; the solid line illustrates the bisector fit.}\label{fig|Rd_Vopt}
\end{figure*}

In Fig.~\ref{fig|Rd_Vopt} we illustrate the relationship between two directly observable quantities, namely the disk scale-length $R_d$ and the optical velocity $V_{\rm opt}$. We plot the datapoints for the individual galaxies of our sample and for the mean within bins of given $R_d$ or of given $V_{\rm opt}$. We also show the linear fits to the mean relationships $R_d-V_{\rm opt}$ and $V_{\rm opt}-R_d$ for binned data, and the associated bisector fit (see Table~\ref{table|fits}). We confront the outcome with the independent measurements by Romanowsky \& Fall (2012) and Lelli et al. (2016), finding a very good agreement. Remarkably the $1\sigma$ dispersion around the mean relation  amounts to about $0.04$ dex.

\begin{figure*}
\epsscale{1}\plotone{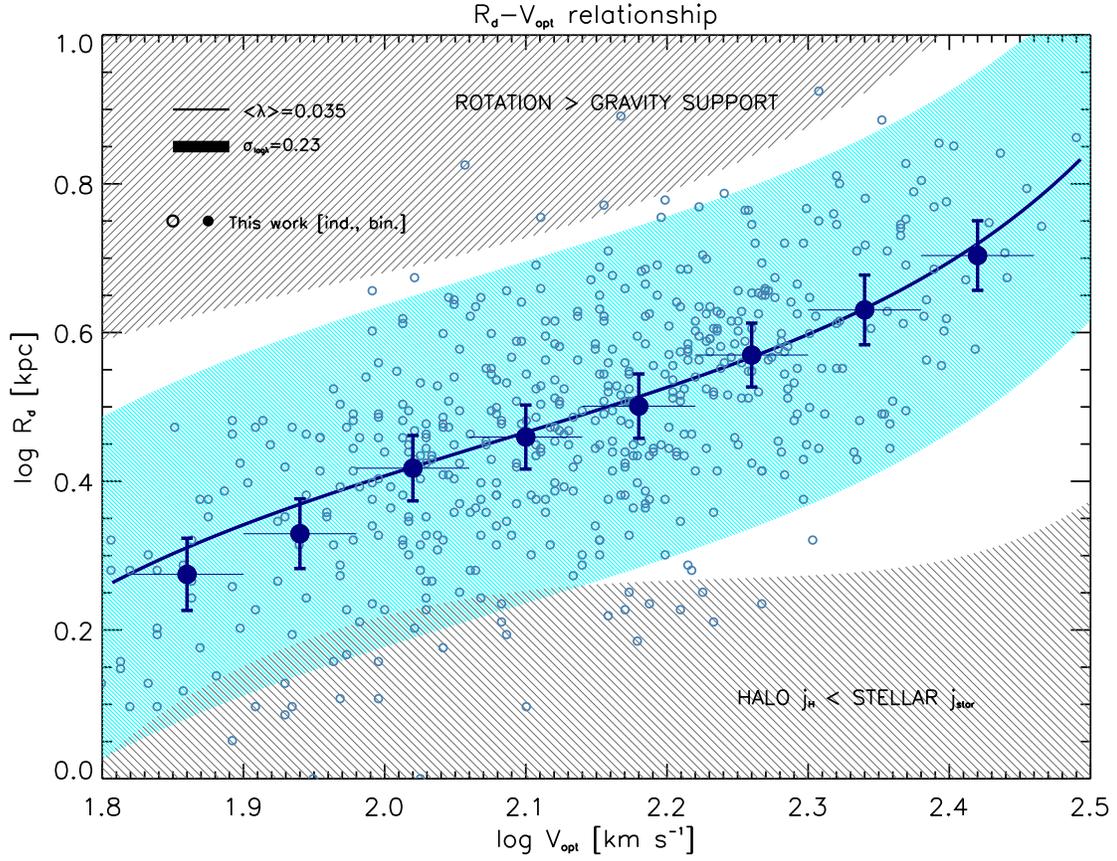}\caption{Same as previous figure, but now the solid line with shaded area is the theoretical relation expected from angular momentum conservation (see Eq.~[\ref{eq|RdVopt}]), with the associated dispersion around $0.25$ dex mainly determined by that in the halo spin parameter; the grey hatched areas illustrate the regions of the diagram where gravitational support is not sufficient to sustain the disk rotational motions (dictated by the halo spin) and where the specific angular momentum of the halo $j_{\rm H}$ would be unphysically lower than that observed in the disk $j_\star$, see Sect.~\ref{sec|RdVopt}.}\label{fig|Rd_Vopt_theory}
\end{figure*}

We now turn to physically interpret this fundamental relation within the classic picture of disk formation, that envisages sharing of specific angular momentum between baryons and DM at halo formation, and then a retention/sampling of almost all momentum into the stellar component. We again equalize $j_\star=f_j\, j_{\rm H}$ with $j_{\rm H}$ expressed as a function of $V_{\rm H}$ from Eq.~(\ref{eq|jhalo}, last term) to $j_\star=2\, f_R\, R_d\, V_{\rm opt}$ from Eq.~(\ref{eq|jstar}). Introducing $f_V\equiv V_{\rm opt}/V_{\rm H}$, we get the straight relation
\begin{equation}\label{eq|RdVopt}
R_d\,[{\rm kpc}]\simeq 1.5\,{\lambda\,f_j\over f_R\,f_V^2}\, E(z)^{-1/2}\, V_{\rm opt}\,[{\rm km~s}^{-1}]~,
\end{equation}
where fits to the dependencies of $f_j$ and $f_V\approx 1.07\,V_{\rm opt}/V_{200}$ on $V_{\rm opt}$ are given in Table~\ref{table|fits}. When adopting the average value of the spin parameter $\langle\lambda\rangle\approx 0.035$ indicated by numerical simulations, the relation $R_d-V_{\rm opt}$ after Eq.~(\ref{eq|RdVopt}) is plotted in Fig.~\ref{fig|Rd_Vopt_theory}; it remarkably agrees in normalization and shape with the observed one. We note that using the simplified expression $R_d\,[{\rm kpc}]\simeq 0.64\,\lambda\, V_{\rm opt}\,[{\rm km~s}^{-1}]$, that involves the mean value $\langle f_j/f_R\,f_V^2\rangle\approx 0.43$ over the different $V_{\rm opt}$ bins, yields in turn a very good representation of the bisector fit relation presented in the previous Fig.~\ref{fig|Rd_Vopt}.

There is an issue, however, that needs to be discussed with some care. The $1\sigma$ variance expected from the theoretical relation of Eq.~(\ref{eq|RdVopt}), which is mainly determined by that in the halo spin parameter $\lambda$, amounts to about $0.25$ dex; this $1\sigma$ dispersion is seen to embrace almost all the data, and indeed it is appreciably larger (by a factor $2$) than the observed $1\sigma$ scatter (see also Saintonge et al. 2008, 2011; Hall et al. 2012). This may be partly due to the fact that stable disks must be hosted by halos with a quiet recent merging histories (see Wechsler et al. 2002; Dutton et al. 2007). In addition, specific regions (highlighted by hatched areas) of the $R_d-V_{\rm opt}$ diagram, that correspond to very high or very low values of $\lambda$  are actually prohibited by simple physical arguments. Specifically, too high values of $\lambda$, that would imply large values of $R_d$ at given $V_{\rm opt}$, are not permitted because otherwise the gravitational support would be not sufficient to sustain the rotational motions of the stars; this condition can be naively expressed as $G\, M_{\rm tot}/R_d^2\sim j_\star^2/R_d^3$, implying $\lambda\la f_b/2\sqrt{2}\, f_V\, f_R$. Contrariwise, too low values of $\lambda$, that would imply small values of the disk scale-length $R_d$ at given $V_{\rm opt}$, are not permitted since otherwise the specific angular momentum $j_{\rm H}\propto \lambda\, V_{\rm H}^2\propto \lambda\, V_{\rm opt}^2$ of the halo (cf. Eq.~\ref{eq|jhalo} and Fig.~\ref{fig|Vratio_Mstar}-\ref{fig|jstar_Mh}) would be smaller than the measured stellar one $j_\star\propto f_j\, \langle\lambda\rangle$; this implies $\lambda\ga f_j\, \langle\lambda\rangle$ with $\langle\lambda\rangle\approx 0.035$. A similar criterion is obtained when basing on theoretical disc instability arguments (see Mo et al. 1998).

The tightness and the theoretical understanding of the mean $R_d-V_{\rm opt}$ relation opens up the possibility of exploiting it in the next future for cosmological studies. First of all, such a relation involves two directly observable quantities, of which $R_d$ is determined from photometry and as such depends on the angular diameter distance, while $V_{\rm opt}$ can be estimated from kinematic measurements and is independent of it (see also Saintonge et al. 2008, 2011). Second, the redshift evolution of the relationship after Eq.~(\ref{eq|RdVopt}) is expected to be mild and controlled, since $\lambda$ is known from simulations to be almost independent of mass and redshift; the other shape factors are weakly varying function of $V_{\rm opt}$ and should be related only to internal structure of pure disks, which is known to be weakly dependent on redshift at least out to $z\la 0.7$ (see Hudson et al. 2015; Huang et al.
2017). Therefore, observations at increasing redshift can in principle be exploited to constrain the Hubble scale and the cosmological parameters, with particular focus on the dark energy properties.

\section{Summary}\label{sec|summary}

We have built templates of rotation curves as a function of luminosity via the mass modeling (by the sum of a thin exponential disk and a cored halo profile) of suitably normalized, stacked data from wide samples of local spiral galaxies (Sect.~\ref{sec|URC}). We have then exploited such templates to determine fundamental stellar and halo properties for a sample of about $550$ local disk-dominated galaxies (Sect.~\ref{sec|sample}) with high-quality measurements of the optical radius $R_{\rm opt}$ and of the corresponding rotation velocity $V_{\rm opt}$. We have determined global quantities such as the stellar $M_\star$ and halo $M_{\rm H}$ masses, the halo size $R_{\rm H}$ and velocity scale $V_{\rm H}$, and the specific angular momenta of the stellar $j_\star$ and halo $j_{\rm H}$ components. The relevant quantities for each galaxy in our sample are summarized in Table~\ref{table|data}.

We have derived global scaling relationships involving such properties, both for the individual galaxies in our sample and for their mean within bins (see Sect.~\ref{sec|results}). Specifically, we have presented the following relations: halo mass $M_{\rm H}$ vs. stellar mass $M_\star$ and related star formation efficiency $f_\star=M_\star/0.16\, M_{\rm H}$, see Sect.~\ref{sec|Mh_Mstar} and Fig.~\ref{fig|Mh_Mstar}; effective radius $R_e\approx 1.68\, R_d$ vs. stellar mass, see Sect.~\ref{sec|sizes} and Fig.~\ref{fig|Re_Mstar}; central surface density $\Sigma_0=M_\star/2\pi\,R_d^2$ vs. stellar mass $M_\star$, see Sect.~\ref{sec|sizes} and Fig.~\ref{fig|Sigma0_Mstar}; stellar mass $M_\star$ and $M/L$ ratio vs. optical velocity $V_{\rm opt}$, see Sect.~\ref{sec|MoverL} and Fig.~\ref{fig|Mstar_Vopt}; the fundamental plane of spirals, see Sect.~\ref{sec|MoverL} and Fig.~\ref{fig|FP}; the ratio $V_{\rm opt}/V_{200}$ between optical velocity $V_{\rm opt}$ to halo velocity $V_{200}\approx 1.07\, V_{\rm H}$ vs. stellar mass $M_\star$ or vs. optical velocity $V_{\rm opt}$, see Sect.~\ref{sec|velratio} and Fig.~\ref{fig|Vratio_Mstar}; specific angular momentum $j_\star$ vs. stellar mass $M_\star$ or vs. optical velocity $V_{\rm opt}$ or vs. halo mass $M_{\rm H}$, see Sect.~\ref{sec|jstar} and Figs.~\ref{fig|jstar_Mstar} and \ref{fig|jstar_Mh}; effective radius $R_e$ vs. halo radius $R_{200}$, see Sect.~\ref{sec|ReR200c} and Fig.~\ref{fig|Re_R200c}; disk scale-length $R_d$ vs. optical velocity $V_{\rm opt}$, see Sect.~\ref{sec|RdVopt} and Fig.~\ref{fig|Rd_Vopt}.

We have found our results to be in pleasing agreement with previous determinations by independent methods like abundance matching, weak lensing observations, and individual RC modeling. We stress that the size of our sample and the robustness of our approach have allowed us to attain an unprecedented level of precision over an extended range of mass and velocity scales, with $1\sigma$ dispersion around the mean relationships of less than $0.1$ dex. These results can provide a benchmark to gauge determinations from independent techniques, such as weak gravitational lensing or abundance matching, and check for systematics.

We have physically interpreted the fundamental $R_d-V_{\rm opt}$ relationship within the classic picture of disk formation, that envisages sharing of specific angular momentum between baryons and DM at halo formation, and then a retention/sampling of almost all angular momentum into the stellar component (see Sect.~\ref{sec|RdVopt} and Fig.~\ref{fig|Rd_Vopt_theory}). Remarkably, the observed $R_d-V_{\rm opt}$ relationship strongly supports such a scenario, and robustly indicate the average value $\langle\lambda\rangle\approx 0.035$ for the halo spin parameter, perfectly in agreement with $N$-body simulations.

Moreover, we have elucidated that the small scatter of the observed $R_d-V_{\rm opt}$ relationship with respect to the theoretical prediction (related to the dispersion in the halo spin parameter $\lambda$) can be explained by requiring sufficient gravitational support against rotation and by having a specific angular momentum of the halo greater than that measured for the disk. Finally, we have stressed that the tightness and the theoretical understanding of the mean $R_d-V_{\rm opt}$ relation opens up the
perspective of exploiting it in the near future for cosmological studies.

Two observational efforts are in order to pursue such a program. First, the RCs for a significant sample of local disk galaxies should be extended out to $R\sim 3\, R_{\rm opt}$ (see, e.g., Martinsson et al. 2016); this will allow to drastically reduce the systematic errors in the estimate of template RCs as a function of luminosity, which constitute the main source of uncertainty in the $R_d-V_{\rm opt}$ relationship. Second, measurement of the disk scale-length and optical velocity for disk-dominated galaxies toward higher redshift $z\la 1$ would be of paramount importance both for tracing the astrophysical evolution of the spiral population and for exploiting it as a cosmological probe.

All in all, our results have set a new standard of precision in the global relationships obeyed by local disk-dominated galaxies, that must be reproduced by detailed physical models, that offers a basis for improving the sub-grid recipes in numerical simulations, that provides a benchmark to gauge independent observations and check for systematics, and that constitutes a basic step toward the future exploitation of the spiral galaxy population as a cosmological probe.

\begin{acknowledgements}
We thank the referee for a constructive report. We are grateful to A. Bressan and F. Fraternali for stimulating discussions. Work partially supported by PRIN MIUR 2015 `Cosmology and Fundamental Physics: illuminating the Dark Universe with Euclid'. AL acknowledges the RADIOFOREGROUNDS grant (COMPET-05-2015, agreement number 687312) of the European Union Horizon 2020 research and innovation programme, and the MIUR grant `Finanziamento annuale individuale attivit\'a base di ricerca'.
\end{acknowledgements}

\begin{appendix}

\section{DM halo profiles}

In this Appendix we discuss the dependence of our results on the adopted DM halo profile. The halo RC, normalized to the value at the optical radius $R_{\rm opt}$, can be written as (see Sect.~\ref{sec|URC})
\begin{equation}\label{eq|app_Vhalo}
{V_{\rm HALO}^2(R)\over V_{\rm HALO}^2(R_{\rm opt})} = {1\over \tilde R}\,{G(\tilde R/\tilde R_c)\over G(1/\tilde R_c)}~,
\end{equation}
with $\tilde R\equiv R/R_{\rm opt}$; the shape factor $G(x)$ depends on the specific halo profile adopted, and $R_c$ is a characteristic radius entering its expression.

In the main text we have adopted as a reference the cored Burkert density profile $\rho(R)\propto (1+R/R_c)^{-1}\,[1+(R/R_c)^2]^{-1}$, with $R_c$ representing the core radius (see Burkert 1995). The associated RC shape factor entering Eq.~(\ref{eq|app_Vhalo}) reads
\begin{equation}
G(x)=\ln(1+x)-{\rm tan}^{-1}(x)+{1\over 2}\,\ln(1+x^2)~.
\end{equation}

Here we consider two alternative choices. The first one is the classic NFW density profile $\rho(R)\propto (R/R_c)^{-1}\,(1+R/R_c)^{-2}$ extracted from $N-$body simulation (e.g., Navarro et al. 1997, 2010; Ludlow et al. 2013; Peirani et al. 2017); here the characteristic radius $R_c$ represents the point where the logarithmic density slope equals $-2$. The associated RC shape factor entering Eq.~(\ref{eq|app_Vhalo}) reads
\begin{equation}
G(x)=\ln(1+x)-{x\over (1+x)}~.
\end{equation}

The second one is the density profile found by hydrodynamical simulations that include baryonic cooling/feedback processes, and their ensuing contraction/expansion effects on the DM halo (e.g., Di Cintio et al. 2014; hereafter DC profile). The outcomes of such simulations can be rendered as a generalized NFW density profile $\rho(R)\propto (R/R_c)^{-\gamma}\,[1+(2-\gamma)\,(R/R_c)^\alpha/(\beta-2)]^{-(\beta-\gamma)/\alpha}$, characterized by three shape parameters $(\alpha,\beta,\gamma)$, and with $R_c$ being again the radius where the logarithmic density slope is $-2$; the NFW profile corresponds to $(\alpha,\beta,\gamma)=(1,3,1)$. Actually the triple $(\alpha,\beta,\gamma)$ depends on the stellar-to-halo mass ratio $M_\star/M_{\rm H}$ (see Di Cintio et al. 2014, their Fig.~1); for the sake of definiteness, here we take the values $(\alpha,\beta,\gamma)=(5/2,5/2,1/2)$ approximately holding for an extended range $M_\star/M_{\rm H}\sim 10^{-3.5}-10^{-1.5}$. The resulting profile strikes an intermediate course between the cored Burkert and the cuspy NFW. Noticeably, the associated RC shape factor can be written in closed analytical form as
\begin{equation}
G(x)=-1+(1+3\, x^{5/2})^{1/5}~.
\end{equation}

\begin{figure*}
\epsscale{1}\plotone{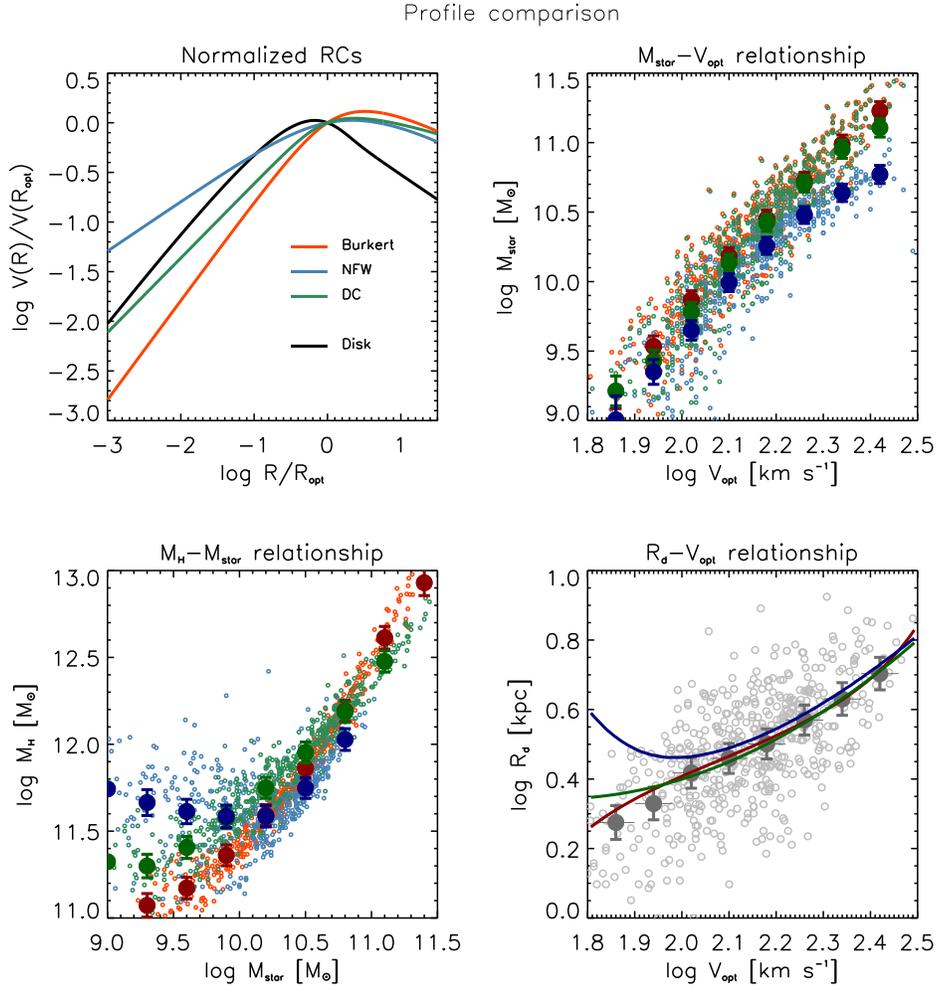}\caption{Dependence of the results from this work on the adopted DM halo profile. Top left panel: halo RCs $V_{\rm HALO}(R)$ normalized to the value $V_{\rm HALO}(R_{\rm opt})$ at the optical radius $R_{\rm opt}$. Top right panel: $M_\star-V_{\rm opt}$ relationship; bottom left panel: $M_{\rm H}-M_\star$ relationship; bottom right panel: $R_d-V_{\rm opt}$ relationship. In all panels red symbols/lines refer to the Burkert profile, blue symbols/lines to the NFW profile, and green symbols/lines to the DC profile (see Appendix for details). In addition, small dots refer to the values inferred for individual galaxies, big circles refer to the mean within bins of the $x-$axis variable, and lines in the $R_d-V_{\rm opt}$ diagram are the theoretical expectations based on Eq.~(\ref{eq|RdVopt}). }\label{fig|app}
\end{figure*}

The normalized halo RCs associated to the Burkert, NFW and DC profiles with $\tilde R_c=R_c/R_{\rm opt}=1$ are illustrated in Fig.~\ref{fig|app}; for reference the disk component is also shown. We have redone the analysis of the main text by using the above profiles; the comparison of the results on the fundamental $M_\star-V_{\rm opt}$, $M_{\rm H}-M_\star$, and $R_d-V_{\rm opt}$ relationships is also reported in Fig.~\ref{fig|app}.

The distinct shape of the halo RCs in the inner radial range $R\la R_{\rm opt}$, where the rather steep disk component can be relevant, causes the estimated stellar masses to be appreciably different for the three halo profiles. Halo RCs with flatter shape in the inner regions (corresponding to cuspier density profiles) require a less prominent disk contribution and hence a smaller $M_\star$. The DC profile yields stellar masses smaller by a factor $1.2-1.5$, and the NFW profile smaller by a factor $1.5-2.5$, than the Burkert profile; at the brightest magnitudes, the stellar masses inferred with the DC profile hardly exceed $10^{11}\, M_\odot$ and with the
NFW profile stay below several $10^{10}\, M_\odot$.

As to the halo masses, for the majority of galaxies (with intermediate and bright magnitudes) the behavior of the RCs in the outer radial range $R\ga R_{\rm opt}$ is the most relevant. Halo RCs increasing for a more extended radial range and/or featuring a flatter decrease beyond the maximum tend to yield larger halo masses; thus the Burkert profile provides the largest halo masses, followed by the DC and the NFW profiles. However, in faint galaxies where the disk contribution is marginally relevant or even negligible, the halo mass is strongly dependent also on the portion of the RC at $R\la R_{\rm opt}$, with flatter shapes (corresponding to cuspier density profiles) yielding larger masses; this is the origin of the flattening, or even upturn, of the $M_{\rm H}-M_\star$ relationship at the small-mass end for the DC and NFW profiles.

Despite these differences, it is remarkable that the theoretically expected $R_d-V_{\rm opt}$ relationship based on Eq.~(\ref{eq|RdVopt}) is very similar for the three halo profiles, and in agreement with the observed mean relation, at least for $V_{\rm opt}\ga 150$ km s$^{-1}$; this is ultimately due to the weak dependence of the quantity $f_j/f_R\,f_V^2$ appearing in Eq.~(\ref{eq|RdVopt}) on the assumed halo profile. As a consequence, the theoretical interpretation of the $R_d-V_{\rm opt}$ relationship appears to be robust against variation in the DM halo profile; this is very encouraging in view of the future exploitation of the relationship for cosmological studies.

\end{appendix}

\clearpage

\clearpage

\clearpage

\clearpage

\clearpage

\clearpage

\clearpage

\clearpage

\clearpage

\clearpage

\clearpage

\clearpage

\clearpage

\clearpage

\clearpage

\clearpage

\clearpage
\begin{deluxetable}{lrrrrrrrrrrrrrrrrrrrrrrrrrrr}
\tablewidth{0pt}\tablecaption{Fits to global relationships for local spiral galaxies}\tablehead{\colhead{Relation} & &\multicolumn{7}{c}{Fitting Parameters}\\
\cline{1-1} \cline{3-9}
\colhead{$y-x$} & &\colhead{$y_0$} & & \colhead{$y_1$} & & \colhead{$y_2$} & &
\colhead{$y_3$}}\startdata
                $\log M_{\rm H}-\log M_\star$ & &    11.8562375    & &     0.9284897    & &     0.2066120    & &     0.0005020     \\
                       $\log R_e-\log M_\star$ & &     0.7519445    & &     0.2332562    & &     0.0494138    & &     0.0267494     \\
                  $\log \Sigma_0-\log M_\star$ & &     8.6510640    & &     0.4887056    & &    -0.1162257    & &    -0.0515896     \\
               $\log M_\star-\log V_{\rm opt}$ & &    10.5269449    & &     3.4242617    & &    -1.2325658    & &     1.3322386     \\
           $\log M_\star/L_I-\log V_{\rm opt}$ & &     0.2377460    & &     0.8276025    & &    -0.3334509    & &     4.9630465     \\
            $V_{\rm opt}/V_{200}-\log M_\star$ & &     1.2525990    & &    -0.1582360    & &    -0.1246982    & &    -0.0099017     \\
        $V_{\rm opt}/V_{200}-\log V_{\rm opt}$ & &     1.2620718    & &    -0.3582905    & &    -1.8746358    & &    -1.0999992     \\
            $V_{\rm opt}/V_{200}-\log R_{200}$ & &     1.2845604    & &    -0.2741447    & &    -1.5439286    & &     1.0877142     \\
\enddata
\tablecomments{Relation $y-x$ is fitted with the polynomial shape $y=y_0+y_1\, (x-\tilde x)+y_2\, (x-\tilde x)^2+y_3\, (x-\tilde x)^3$. Masses are in solar units, sizes in kpc, velocities in km s$^{-1}$, and specific angular momenta in km s$^{-1}$ kpc. Normalization points are chosen as $\log \tilde M_\star = 10.5$, $\log \tilde V_{\rm opt} = 2.2$, $\log \tilde R_{200} = 2.2$, $\log \tilde R_{d} = 0.5$, and $\log \tilde M_{\rm H} = 12.0$.
Fits hold to within $5\%$ within the ranges: $\log M_\star\sim[9.0,11.5]$, $\log V_{\rm opt}\sim[1.8,2.5]$, $\log R_{200}$ and $\log R_{d}\sim[1.9,2.5]$, and $\log M_{\rm H}\sim[11.0,12.5]$.}\label{table|fits}
\end{deluxetable}

\clearpage
\LongTables
\begin{deluxetable}{lcccccccccccccccc}
\tablewidth{0pt}\tablecaption{Data sample and relevant quantities}\tablehead{\colhead{ESO} & &\colhead{$M_I$} & & \colhead{$\log R_d$} & & \colhead{$\log V_{\rm opt}$} & &
\colhead{$\log M_\star$} & & \colhead{$\log j_\star$} & & \colhead{$\log M_{\rm H}$}\\
\colhead{id} & &\colhead{[mag]} & & \colhead{[kpc]} & & \colhead{[km s$^{-1}$]} & & \colhead{[$M_\odot$]} & & \colhead{[km s$^{-1}$ kpc]} & & \colhead{[$M_\odot$]}}\startdata
546-g36    & &    -18.48    & &      0.20    & &      1.84    & &      7.89    & &      2.28    & &     10.93     \\
291-g24    & &    -18.49    & &      0.14    & &      1.88    & &      7.97    & &      2.26    & &     10.97     \\
602-g15    & &    -18.54    & &      0.30    & &      1.83    & &      8.25    & &      2.38    & &     10.97     \\
357-g16    & &    -18.55    & &      0.05    & &      1.89    & &      8.15    & &      2.19    & &     10.93     \\
437-g47    & &    -18.59    & &      0.29    & &      1.86    & &      8.43    & &      2.39    & &     11.01     \\
60-g25    & &    -18.61    & &      0.07    & &      1.61    & &      7.77    & &      1.93    & &     10.30     \\
441-g11    & &    -18.64    & &      0.28    & &      1.84    & &      8.49    & &      2.36    & &     10.96     \\
551-g31    & &    -18.70    & &      0.15    & &      1.81    & &      8.41    & &      2.21    & &     10.79     \\
\enddata
\tablecomments{Galaxies are sorted in order of brightening $I-$band magnitudes. Typical statistical uncertainties on the tabulated quantities are of order: $0.1$ dex on $\log R_{d}$, $0.03$ dex on $\log V_{\rm opt}$, $0.12$ dex on $\log M_\star$, $0.1$ dex on $\log j_\star$, and $0.15$ dex on $\log M_{\rm H}$.}\label{table|data}
\end{deluxetable}

\end{document}